\documentclass[12pt,english,floatfix,nofootinbib,superscriptaddress,aps,prd,preprint]{revtex4}
\usepackage[utf8]{inputenc}
\usepackage{float}
\usepackage{array}
\usepackage{lipsum}
\usepackage{dsfont}
\usepackage{commath}
\usepackage{graphicx}
\usepackage{amsmath,amsthm,amsfonts,amssymb}
\usepackage{graphicx}
\usepackage[english]{babel} 
\usepackage{color}
\usepackage{tensor}
\usepackage{esint}
\usepackage[dvips]{epsfig}
\usepackage[dvips]{graphicx}
\usepackage{float}
\usepackage{units}
\usepackage{textcomp}
\usepackage{mathrsfs}
\usepackage{amsmath}
\usepackage[makeroom]{cancel}
\usepackage{amssymb}
\usepackage{amsbsy}
\usepackage{amsfonts}
\usepackage{amssymb,mathrsfs,xcolor}
\usepackage{esint}
\usepackage{braket}
\usepackage{array}
\usepackage{graphicx}
\usepackage{caption}
\usepackage{subcaption}

\usepackage{wasysym}
\usepackage{multirow}
\usepackage{wrapfig}

\usepackage{stmaryrd}
\usepackage{upgreek}

\makeatletter

\makeatletter\usepackage{babel}

\usepackage{hyperref}
\hypersetup{
    colorlinks,
    citecolor=blue,
    filecolor=green,
    linkcolor=purple,
    urlcolor=red,
}

\usepackage{slashed}

\newcommand{\ie}{\begin{equation}}
\newcommand{\fe}{\end{equation}}
\newcommand{\se}{\begin{eqnarray}}
\newcommand{\ff}{\end{eqnarray}}

\begin{document}

\title{Thermodynamical properties of an ideal gas in a traversable wormhole}


\author{A. A. Ara\'{u}jo Filho}
\email{dilto@fisica.ufc.br}

\affiliation{Departamento de Física Teórica and IFIC, Centro Mixto Universidad de Valencia--CSIC. Universidad
de Valencia, Burjassot--46100, Valencia, Spain}

\affiliation{Departamento de Física, Universidade Federal da Paraíba, Caixa Postal 5008, 58051--970, João Pessoa, Paraíba, Brazil}

\author{J. Furtado}
\email{job.furtado@ufca.edu.br}

\affiliation{Universidade Federal do Cariri, Centro de Ciências e Tecnologia, 63048-080, Juazeiro do Norte, CE, Brazil}

\author{J. A. A. S. Reis}
\email{jalfieres@gmail.com}

\affiliation{Universidade Estadual do Sudoeste da Bahia (UESB), Departamento de Ciências Exatas e Naturais, Campus Juvino Oliveira, Itapetinga -- BA, 45700-00,--Brazil}

\author{J. E. G. Silva}
\email{euclides.silva@ufca.edu.br}

\affiliation{Universidade Federal do Cariri, Centro de Ciências e Tecnologia, 63048--080, Juazeiro do Norte, CE, Brazil}


\date{\today}

\begin{abstract}

In this work, we analyze the thermodynamic properties of non--interacting particles under influence of the gravitational field of a traversable wormhole. In particular, we investigate how the thermodynamic quantities are affected by the Ellis wormhole geometry, considering three different regions to our study: asymptotically far, close to the throat, and at the throat. The thermodynamic quantities turn out to depend strongly on parameter that controls the wormhole throat radius. By varying it, there exist an expressive modification in the thermodynamic state quantities, exhibiting both usual matter and dark energy--like behaviors. Finally, the interactions are regarded to the energy density and it seems to indicate that it ``cures" the dark energy--like features.


\end{abstract}

\maketitle


\section{Introduction}

General relativity and other extensions of gravitational theories there can accommodate a particular spacetime, possessing a non--trivial topology: the wormhole. In this scenario, it can be seen as a tunnel--like structure, connecting either two different universes or two asymptotically flat regions in the same spacetime. The wormholes arouse naturally as a solution of Einstein's field equations, which was first proposed by Flamm \cite{flamm1916beitrage}, Einstein and Rosen \cite{einstein1935particle}. These studies served as a basis to Wheeler address afterwards some additional consequences in such a context \cite{wheeler1955geons}. At this time, the wormholes were believed to be a mathematical prediction only, since they do not account for traversable spacetimes.

Having a ghost--like scalar field, a new spherically symmetric solution of the wormhole was brought about many years latter by Ellis and Bronnikov \cite{ellis1973ether,bronnikov1973scalar}. After that, Morris and Thorne demonstrated that this class of wormholes was actually traversable \cite{morris1988wormholes}. This one have no horizon and they are singularity--free so that their corresponding tidal effects are so weak.

Although there are several investigations addressing instabilities to the Ellis wormhole \cite{shinkai2002fate,gonzalez2008instability,tsukamoto2015high,bronnikov2012instabilities,shinkai2015wormhole}, there exists on the other hand an exotic matter source that could properly support such a geometry in general relativity (GR) \cite{das2005ellis}. Furthermore, if the modified theories of gravity are taken into account, the ghost--like scalar fields are no longer needed \cite{11,12,13,14,18,19,20,23,24,25}, which makes this traversable wormhole a viable subject for theoretical and observational physics.

Studies have focused on the feasibility of stable, traversable wormholes, often invoking exotic matter for stabilization \cite{visser1995lorentzian}. In addition to potential applications for faster--than--light travel, the mathematical properties of wormholes have been explored, particularly in relation to energy conditions \cite{bouhmadi2014wormholes}. Moreover, some researchers are investigating the cosmological implications of wormholes, such as their role in the early universe or their impact on the formation of cosmic structures \cite{cataldo2009evolving} and \textit{quasinormal} modes \cite{b1,b2,b3,b4,b5}.

In addition, the thermal aspects of a given system are certainly remarkable features worthy to be explored. 
Based exclusively on the modified dispersion relations, the investigation of many facets of a given theory may be undertaken \cite{amelino2001testable}, including its thermal behavior. In the literature, there exists a variety of recent works involving the thermodynamic properties in many different contexts, namely, Lorentz violation \cite{29,30,31,32,33,34,35,36,colladay2004statistical,anacleto2018lorentz,aguirre2021lorentz}, bouncing universe \cite{araujo2021bouncing}, loop quantum gravity \cite{aa2022particles}, Einstein-aether theory \cite{aaa2021thermodynamics}, graviton \cite{aa2021lorentz}, five-dimensional Chern-Simons theory \cite{assunccao2021nonanalyticity} and others \cite{araujo2023thermodynamics,oliveira2019thermodynamic,oliveira2020relativistic,oliveira2020thermodynamic}. Nevertheless, from a modified dispersion relation, there does not exist up to now a thermal study of massive particles in the Ellis wormhole scenario.

In this sense, we provide the calculation to the thermodynamic properties of a gas composed of non-interacting particles living in the Ellis wormhole geometry. For doing so, we use a modified dispersion relation in order to derive all quantities of interest, namely, pressure, internal energy, entropy, and heat capacity. In addition, a specific regime of temperature is considered: the inflationary era ($T = 10^{13}$ GeV).

This paper is organized as follows: in section \ref{II}, we present the particle motion on the Ellis wormhole spacetime and we obtain the modified dispersion relation for our system. In section \ref{III}, we calculate all thermodynamic quantities of interest and we discuss them in three particular regions, i.e., asymptotically far, close to the throat, and at the throat. Finally, in section \ref{IV}, we outline our conclusions.

\section{Particle motion on the Ellis wormhole spacetime} \label{II}

\begin{figure}
    \centering
    \includegraphics[scale=0.55]{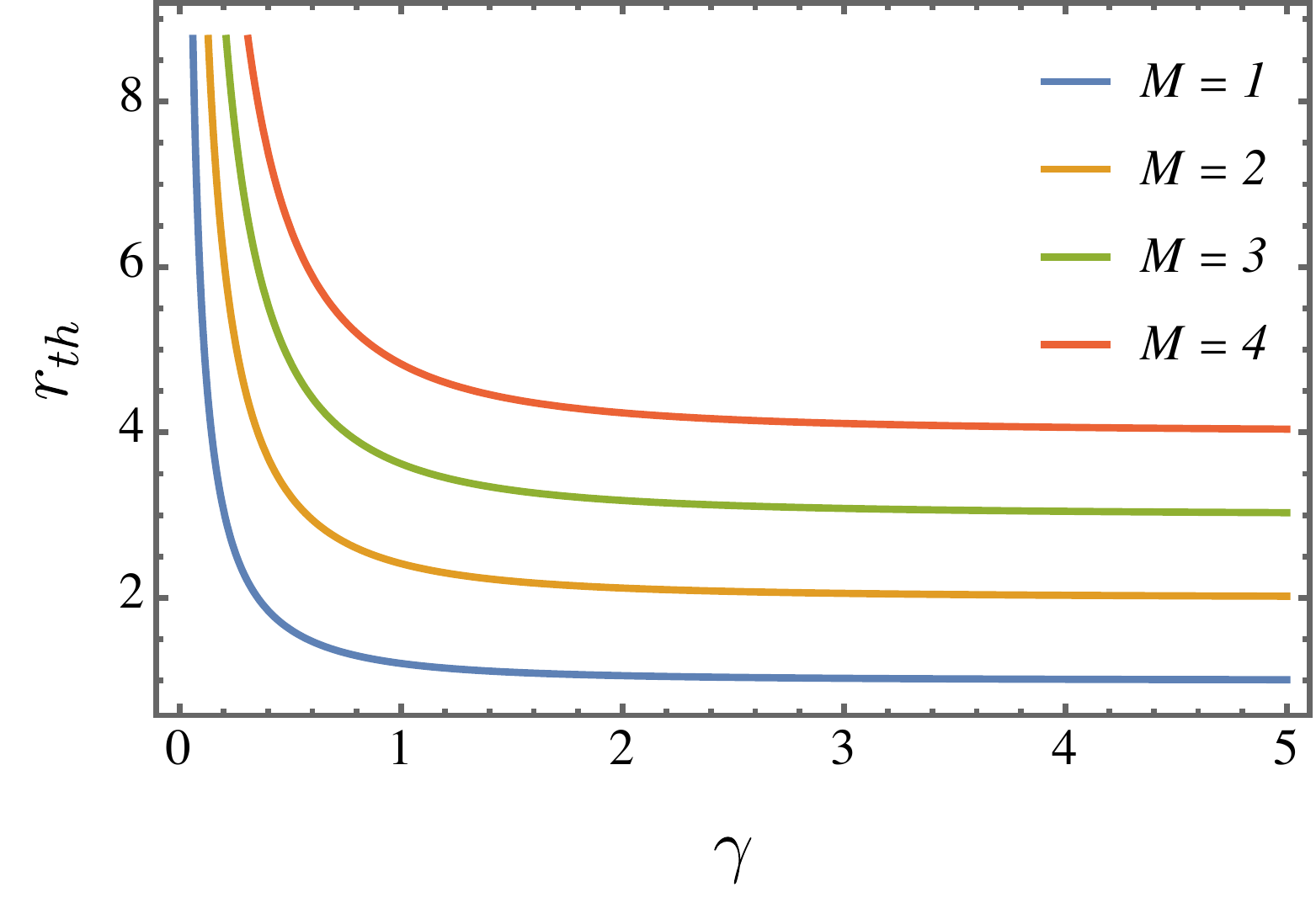}
    \caption{Radius of the wormhole throat as a function of $\gamma$.}
    \label{radius}
\end{figure}

In this section, we shall provide a brief review on the main features of the Ellis wormhole geometry. After that, we derive the relation between Hamiltonian and momentum for a massive particle in such a context. As it is well--known, the Ellis wormhole is a solution of Einstein's equation for a minimally coupled massless scalar field, assuming the following form
\ie
R_{\mu\nu} = \epsilon  \partial_{\mu}\phi \partial_{\nu}\phi,
\fe
where $\phi$ is the scalar field. The ghost regime is represented whether we choose the following configuration $\epsilon = -1$. In the isotropic coordinates, the static spherically symmetric massive Ellis wormhole line element has the form 
\ie
\mathrm{d}\mathrm{s}^{2} = \Omega^{2}(r) \,\mathrm{d}t^{2} - \Phi^{-2}(r)[\mathrm{d}r^{2} + r^{2}(\mathrm{d}\theta + \sin^{2}\theta \,\mathrm{d}\varphi) ], \label{metric}
\fe
where the redshift function $\Omega$ is given by
\ie
\Omega^{2}(r) = e^{2[\epsilon + 2\gamma \tan^{-1}(r/r_{th})]},\label{omegazao}
\fe
$\gamma=-\frac{\epsilon}{\pi}$ and $r_{th}$ is the radius of the wormhole throat.
The isotropic function $\Phi$ reads
\ie
\Phi^{-2}(r) = \left( 1 + \frac{r_{th}^{2}}{4r^{2}} \right)^{2} e^{2[\zeta - 2\gamma \tan^{-1}(r/r_{th})]},\label{phizao}
\fe
and the scalar field has the following solution
\ie
\phi(r) = 4\lambda \tan^{-1}\left(\frac{r}{r_{th}}\right),
\fe
where $2\lambda^2 = 1+\gamma^2$ \cite{nandi2016stability}. 
For $\zeta=\pi\gamma$, the Ellis wormhole is asymptotically flat, i.e., for $r\rightarrow \infty$, $\Omega$, $\Phi$ $\rightarrow 1$. In addition, the scalar field yields to a smooth and complete spacetime without horizon. Additionally, the wormhole radius $r_{th}$ is related to the wormhole mass $M$ and parameter $\gamma$ by \cite{nandi2016stability} 
\ie
r_{th}=\frac{(\gamma + \sqrt{1+\gamma^2})}{2\gamma}M.
\fe
Note that when $\gamma\rightarrow \infty$, $r_{th}$ converges to the constant value, namely, $r_{th}= M$. On the other hand, if $\gamma\rightarrow 0$, the throat radius $r_{th}$ diverges. Thus, parameter $\gamma$ controls the value of the wormhole radius, as shown in Fig. \ref{radius}.


Once described the wormhole geometry, let us study the motion of particles within this complete spacetime. The action for a massive particle reads
\ie
S= -m_0 c \int \mathrm{d}{s},
\fe
where the integral is taken along the particle worldline. Using the metric coefficients from Eq.(\ref{metric}), the particle Lagrangian has the form
\ie
\mathcal{L} \equiv - \ m_{0} c^{2} \Omega \sqrt{ 1 - \frac{v^{2}}{c^{2}}(\Omega\Phi)^{-2} }.
\fe
Analogously, its corresponding momentum in gravitational field is written as
\ie
{\bf{p}} = \frac{ (\Omega\Phi^2)^{-1} m_{0} {\bf{v}}}{ \sqrt{1-\frac{v^{2}}{c^{2}}(\Omega\Phi)^{-2} }} ,
\fe
which leads to the following Hamiltonian
\ie
\mathcal{H} =  m_{0} c^{2}\Omega \sqrt{1 + \Phi^{2}\frac{p^{2}}{m_{0}^{2}c^{2}}}.
\fe
From above equation, de Broglie matter waves can also be derived:
\ie
\label{energymomentumrelation}
\hbar^{2}E^{2} = m^{2}_{0}c^{4} \Omega^{2}\left(1 + \frac{\Phi^2 \hbar^{2}{\bf{k}}^{2}}{(m_0 c)^{2}}\right),
\fe
where, $E$ is the energy, ${\bf{k}}$ is the momentum, and $m_{0}$ is the rest mass. It is worthwhile to mention that, asymptotically, the dispersion relation in Eq.(\ref{energymomentumrelation}) takes the flat form. It is worth highlighting that the modified dispersion relations (MDRs) have emerged as a focal area of research at the interface of quantum mechanics and general relativity \cite{amelino2013quantum,ling2006modified}. The main objective behind them is to address and reconcile the inconsistencies that arise when quantum scale effects interact with gravitational phenomena \cite{smolin2006case,kowalski2005introduction}. In this context, MDRs may provide a clearer understanding of spacetime properties at the Planck scale \cite{mattingly2005modern,girelli2009emergence}.

Moreover, from an observational viewpoint, they could have implications on high--energy processes. There exists a potential for it to influence events such as gamma--ray bursts or the propagation of ultra--high energy cosmic rays \cite{jacob2008lorentz,galaverni2008lorentz}, lending empirical support to theoretical propositions.

In the next section, we examine how the thermodynamic states quantities for a gas of non--interacting particle are modified due to the spacetime curvature of the Ellis wormhole. Since the geometry depends on the wormhole radius and parameter $\gamma$, we consider the variation of the thermodynamics quantities with respect to $r$ and $\gamma$.





\section{Thermodynamic properties} \label{III}

This section is devoted to study the thermodynamic behavior particles from the Broglie matter waves in the context of Ellis wormhole. To do so, we start with the following dispersion relation:  
\ie
\label{energymomentumrelation1}
\hbar^{2}E^{2} = m^{2}_{0}c^{4} \Omega^{2}\left(1 + \frac{\Phi^2 \hbar^{2}{\bf{k}}^{2}}{(m_0 c)^{2}}\right). \nonumber
\fe

In addition, within the framework of MDRs, the exploration of thermodynamic properties holds critical importance \cite{amelino2013quantum}. By astutely understanding the repercussions of MDRs on thermodynamics, we are better positioned to identify and interpret unique patterns in astrophysical observations. This encompasses anomalies in gamma--ray burst time delays \cite{jacob2008lorentz}, ultra--high energy cosmic ray spectra \cite{anchordoqui2003ultrahigh}, or even potential deviations in the radiation profiles of active galactic nuclei \cite{anchordoqui2003ultrahigh}. More so, in Ref. \cite{aaa33}, the authors addressed a similar study considering the \textit{rainbow} gravity instead.

As one could expect, in Eq. (\ref{energymomentumrelation}) a relation between energy and momentum has a different form if compared to the photon--like particle modes. In essence, this aspect gives rise to some interesting remarks that are going to be shown as follows. It is important to mention that, hereafter, we use the natural units to carry out our calculations. From above expression, we may certainly acquire two distinct solutions. However, only one of them agrees with our purpose of possessing real positive definite values:
\begin{eqnarray}
\nonumber{\bf{k}}&=&\frac{1}{c_0 \hbar}\left(\frac{B^2}{r^2}+1\right) e^{\frac{1}{2} \left(-8 \gamma  \tan ^{-1}\left(\frac{r}{B}\right)+2 \zeta -2 \epsilon \right)}\\
&&\times\sqrt{E^2 \hbar ^2-c_0^4 m_0^2 e^{4 \gamma  \tan ^{-1}\left(\frac{r}{B}\right)+2 \epsilon }},
\end{eqnarray}
and, we can straightforwardly write its infinitesimal version $\mathrm{d} {\bf{k}}$ as:
\ie
\mathrm{d} {\bf{k}} = \frac{E \hbar  \left(\frac{B^2}{r^2}+1\right) e^{\frac{1}{2} \left(-8 \gamma  \tan ^{-1}\left(\frac{r}{B}\right)+2 \zeta -2 \epsilon \right)}}{c_0 \sqrt{E^2 \hbar ^2-c_0^4 m_0^2 e^{4 \gamma  \tan ^{-1}\left(\frac{r}{B}\right)+2 \epsilon }}}.
\label{vol}
\fe
Now, we can accomplish the integration over momenta space for the sake of obtaining the accessible states of the system
\ie
\Xi(E) = \int_{0}^{\infty} {\bf{k}}^{2}\,\mathrm{d}{\bf{k}}, \label{ms2}
\fe
where $\Gamma$ is the volume of the thermal reservoir. In this sense, Eq. (\ref{ms2}) reads
\begin{widetext}
\ie
\begin{split}
\Xi(r,\gamma) & =  \int_{0}^{\infty} \frac{E}{c_0^3 \hbar }  \left(\frac{B^2}{r^2}+1\right)^3  e^{\left(\frac{1}{2} \left(-8 \gamma  \tan ^{-1}\left(\frac{r}{B}\right)+2 \zeta -2 \epsilon \right)-8 \gamma  \tan ^{-1}\left(\frac{r}{B}\right)+2 \zeta -2 \epsilon \right)} \\
& \times \sqrt{E^2 \hbar ^2-c_0^4 m_0^2 e^{4 \gamma  \tan ^{-1}\left(\frac{r}{B}\right)+2 \epsilon }} \,\mathrm{d}E.
\end{split}
\fe
\end{widetext}
It is worth mentioning that all calculations performed in this manuscript will be provided in a ``per volume'' manner. In order to supply a better comprehension to the reader, we write the most general definition of the partition function \cite{greiner2012thermodynamics}: 
\ie
Z = \frac{1}{N!h^{3N}} \int \mathrm{d}q^{3N}\mathrm{d}k^{3N} e^{-\beta H(q,p)}  \equiv \int \mathrm{d}E \,\Xi(E) e^{-\beta E}, \label{partti1}
\fe
which accounts for an indistinguishable spinless gas. Moreover, we have considered that $h$ is the Planck's constant, $\kappa_{B}$ is the Boltzmann constant, $\beta = 1/\kappa_{B}T$, $k$ is the generalized momenta, $q$ is the generalized coordinates, $H(k,q)$ is the Hamiltonian of the system, and $N$ is the number of particles. Nevertheless, we see that Eq. (\ref{partti1}) does not address the respective spin of particles under consideration. To step towards, such a feature must be introduced as follows \cite{ghosh1,ghosh2,hassanabadi2022thermodynamics} 
\ie
\mathrm{ln}[Z] = \int \mathrm{d}E \,\Xi(E) \mathrm{ln} [ 1- e^{-\beta E}],
\fe
where $\mathrm{ln} [ 1- e^{-\beta E}]$ accounts for bosons through the Bose-Einstein distribution. Thereby, we can write the partition function as
\begin{widetext}
\ie\label{lnz1}
\begin{split}
\mathrm{ln}[Z(r,\gamma)] & = - \int_{0}^{\infty} \frac{E}{c_0^3 \hbar }  \left(\frac{B^2}{r^2}+1\right)^3  e^{\left(\frac{1}{2} \left(-8 \gamma  \tan ^{-1}\left(\frac{r}{B}\right)+2 \zeta -2 \epsilon \right)-8 \gamma  \tan ^{-1}\left(\frac{r}{B}\right)+2 \zeta -2 \epsilon \right)} \\
& \times \sqrt{E^2 \hbar ^2-c_0^4 m_0^2 e^{4 \gamma  \tan ^{-1}\left(\frac{r}{B}\right)+2 \epsilon }}\, \mathrm{ln} [ 1- e^{-\beta E}] \mathrm{d}E.
\end{split}
\fe
\end{widetext}
With above expression, all thermal quantities of interest can properly be investigate in the next sections. As we shall verify, the expressions does not have \textit{analytical} solutions. Rather, we investigate them in a \textit{numerical} manner. We exhibit the definitions of the thermodynamic functions:
\ie
\begin{split}
  & P (\beta,r,\gamma)= \frac{1}{\beta} \mathrm{ln}\left[Z(\beta,r,\gamma)\right], \\
 & U(\beta,r,\gamma)=-\frac{\partial}{\partial\beta} \mathrm{ln}\left[Z(\beta,r,\gamma)\right], \\
 & S(\beta,r,\gamma)=k_B\beta^2\frac{\partial}{\partial\beta}F(\beta,r,\gamma), \\
 & C(\beta,r,\gamma)=-k_B\beta^2\frac{\partial}{\partial\beta}U(\beta,r,\gamma),
\label{properties}
\end{split}
\fe  
where $P(\beta,r,R,\gamma)$ is the pressure, $U(\beta,r,R,\gamma)$ is the mean energy, $S(\beta,r,R,\gamma)$ is the entropy, and $C(\beta,r,R,\gamma)$ is the heat capacity. In the following subsection, we start examining the equation of states.


\subsection{Pressure}

Here, let us start exploring the effects due to the geometry on the equation of states of the system. In this manner, the pressure can straightforwardly be calculated, giving us
\begin{widetext}
\ie
\begin{split}
P(\beta,r,\gamma)  & = - \int_{0}^{\infty} \frac{E}{\beta c_0^3 \hbar }  \left(\frac{B^2}{r^2}+1\right)^3  e^{\left(\frac{1}{2} \left(-8 \gamma  \tan ^{-1}\left(\frac{r}{B}\right)+2 \zeta -2 \epsilon \right)-8 \gamma  \tan ^{-1}\left(\frac{r}{B}\right)+2 \zeta -2 \epsilon \right)} \\
& \times \sqrt{E^2 \hbar ^2-c_0^4 m_0^2 e^{4 \gamma  \tan ^{-1}\left(\frac{r}{B}\right)+2 \epsilon }}\, \mathrm{ln} [ 1- e^{-\beta E}] \mathrm{d}E.
\end{split}
\fe
\end{widetext}
Next, we shall analyze this thermodynamic equation  for three different configurations, involving the radius $r$, namely, an asymptotic behavior, a point at the throat, and near to it.

\subsubsection{Asymptotic behavior}

Here, we have to consider the limit where $r \rightarrow \infty$. In this case, $\epsilon = - \gamma \pi$, and $\zeta = \gamma \pi$. Thereby, the pressure $P_a(\beta,\gamma)=\lim_{r \to \infty} P(\beta,r,\gamma)$ will be discussed numerically. 
\begin{figure}[tbh]
  \centering
\includegraphics[width=8cm,height=5cm]{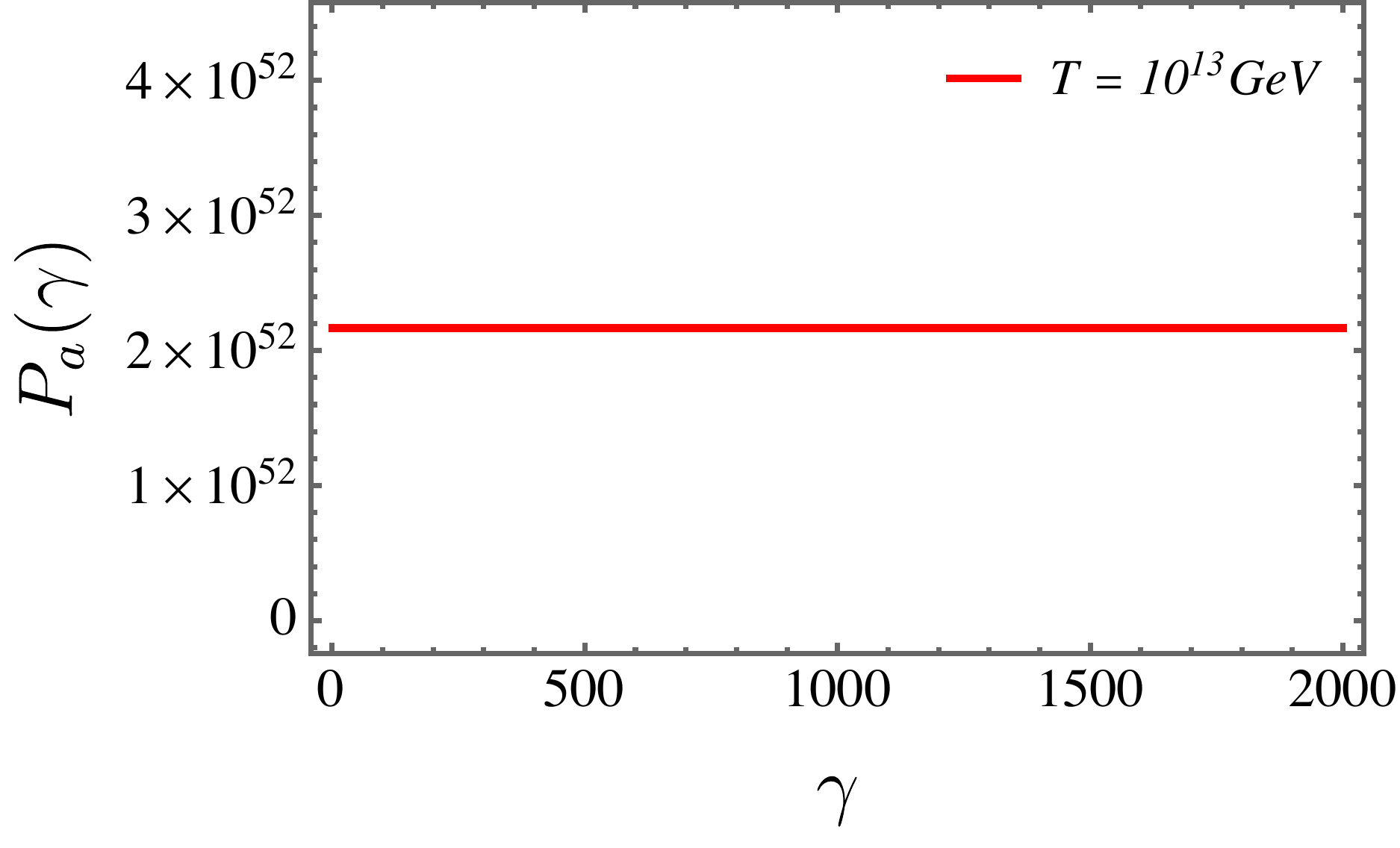}
  \caption{Pressure considering the asymptotic region of the spacetime, namely, $r \rightarrow \infty$ 
  for the inflationary era regime $(T=10^{13}$ GeV) }\label{Equation of states-asymp}
\end{figure}
In Fig. \ref{Equation of states-asymp}, we have displayed the pressure behavior for a particular regime of temperature of the universe: the inflationary era, i.e., $T=10^{13}$ GeV. As it can be seen, $P_{a}(\gamma)$ assumes the same constant value. This feature reflects the fact that, far from the wormhole throat, the spacetime is asymptotically flat and the gas behaves like an usual gas composed by non--interacting scalar particles at equilibrium.


\subsubsection{At the throat}

Now, we analyze how pressure behaves whether we consider a point located at the wormhole throat, namely $r=r_{th}= M(\gamma + \sqrt{1+\gamma^{2}})/2\gamma$. 
\begin{figure}[tbh]
  \centering
  \includegraphics[width=8cm,height=5cm]{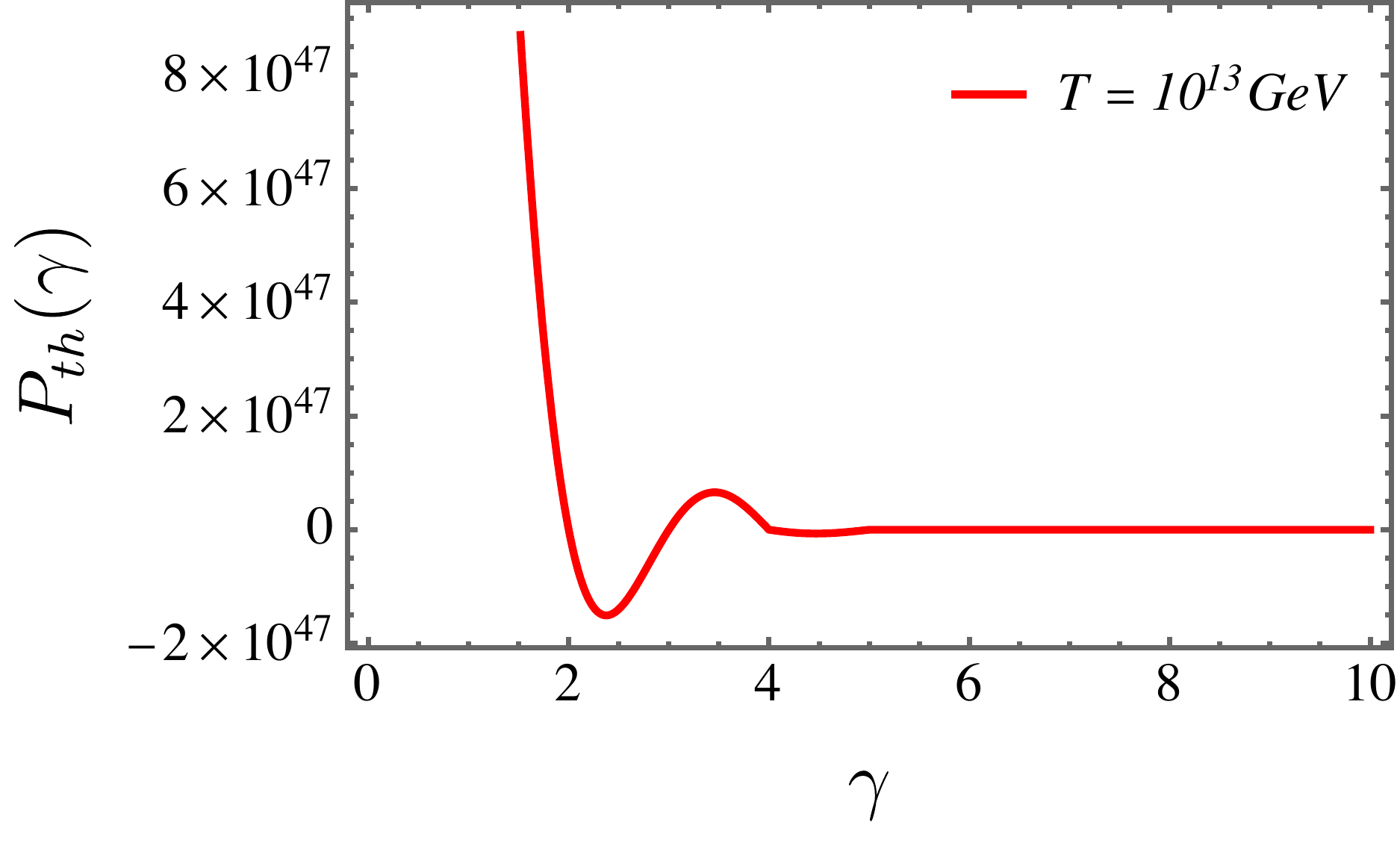}
  \caption{Pressure considering the region on the throat, i.e., $r=r_{th}$, 
  for the inflationary era $(T=10^{13}$ GeV)}\label{Equation of states-throat}
\end{figure}
Remarkably, we have an interesting behavior of pressure $P_{th}(\gamma)$, as displayed in Fig \ref{Equation of states-throat}. Such a plot indicates a dark matter--like behavior when a particular range of parameter $\gamma$ is taken into account, namely, $2<\gamma<3$. In this case, the pressure at the throat becomes negative. Note that, in a general panorama, of $P_{th}(\gamma)$ resembles the behavior of throat radius encountered in Fig. \ref{radius}.


\subsubsection{Near to the throat}

Unless stated otherwise, in order to study how the thermodynamic functions are modified when we regard regions near to the throat, we choose $r=M\sqrt{1+\gamma^{2}}/2 \gamma$ hearafter.

\begin{figure*}[ht]
\centering
\begin{subfigure}{.4\textwidth}
  \centering
  \includegraphics[scale=0.4]{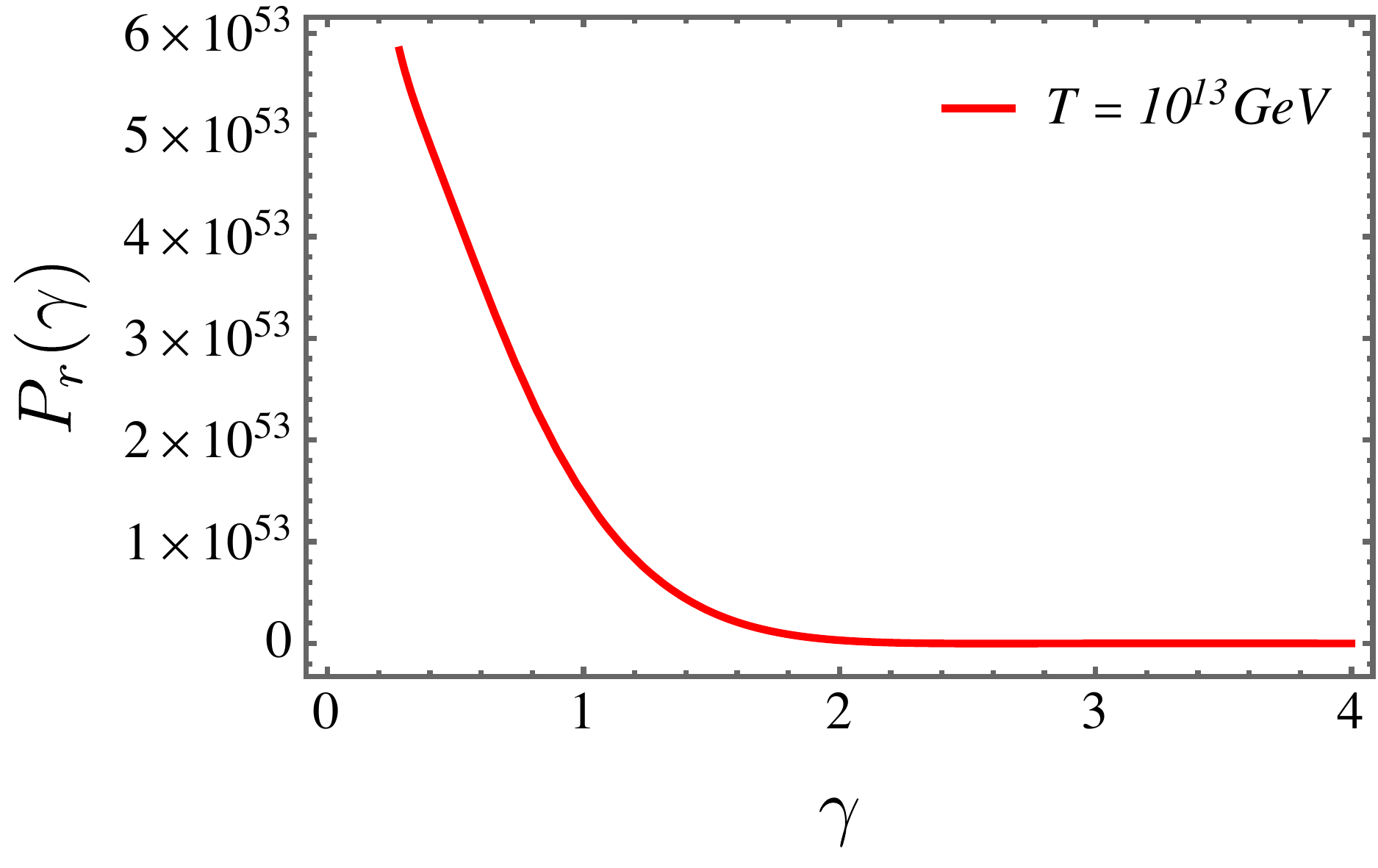}
  \caption{}
\end{subfigure}%
\begin{subfigure}{.6\textwidth}
  \centering
  \includegraphics[scale=0.4]{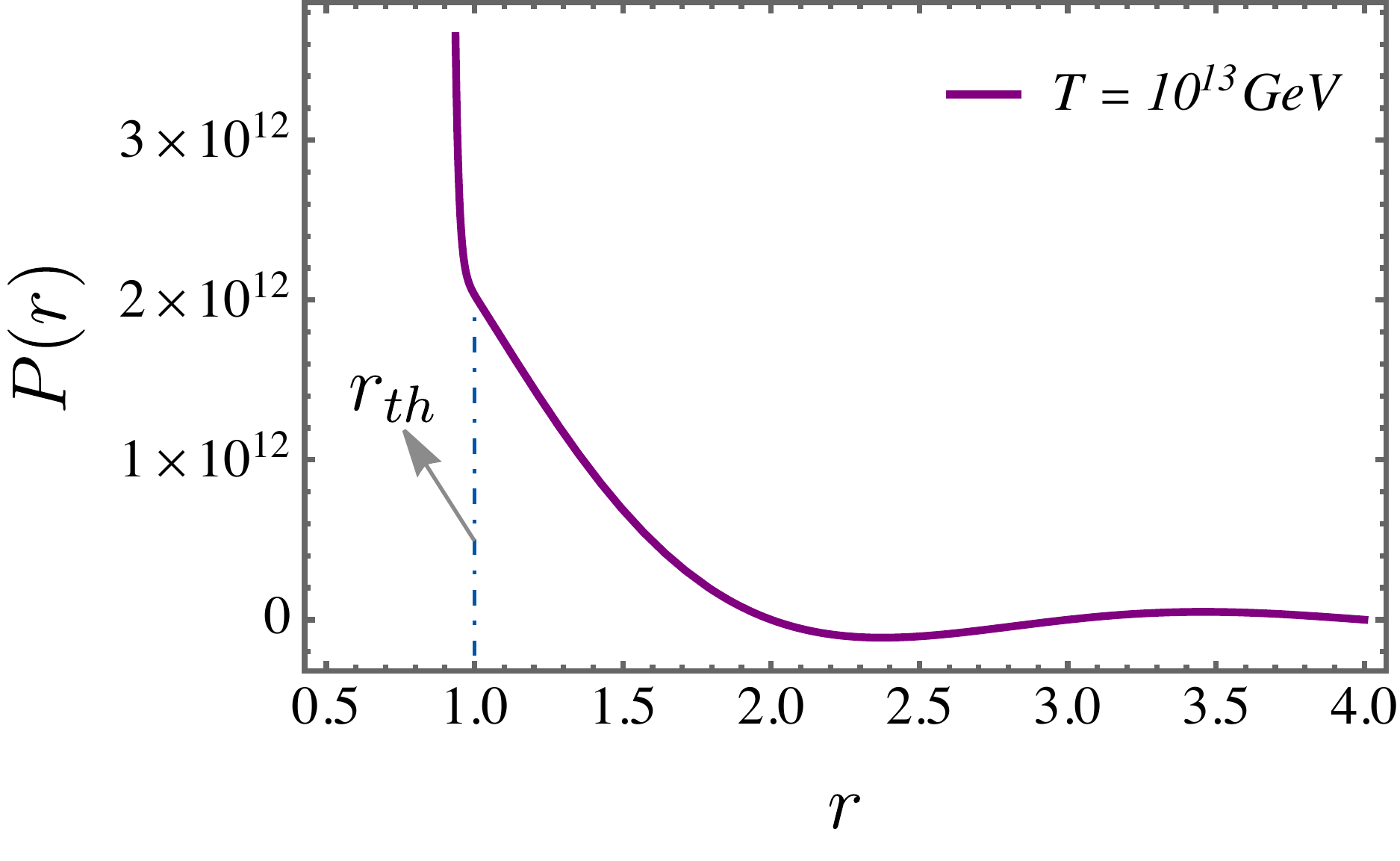}
  \caption{}
 
\end{subfigure}%
\caption{Pressure considering the region near to the throat, i.e., $r=M\sqrt{1+\gamma^{2}}/2 \gamma$, 
  in the inflationary era $(T=10^{13}$ GeV). The panel (a) displays the pressure as a function of $\gamma$ while the panel (b) displays the pressure as a function of $r$.}\label{Equation of states-near-throat}
\end{figure*}


In Fig. \ref{Equation of states-near-throat}, we verify that pressure $P_{r}(\gamma)$, which in this case it is referred to lie near to wormhole throat, tends to attenuate its magnitude as we increase the values of $\gamma$. In addition, to provide a concise investigation in this scenario, we have studied such a behavior as a function of the radial coordinate $r$ as well. Next, we can see in Fig. \ref{Equation of states-near-throat} that there exists a particular region close to the wormhole throat ($2<r<3$), which there may also accommodate dark matter--like behavior of a gas with negative values of pressure. For this plot, we have considered such a normalization\footnote{It is worthy to be highlighted that the same normalization will be regarded to the other thermodynamic properties in this manuscript.}: $r_{th}=1$.

In addition, another aspect worthy to be investigated is the analysis of the pressure as a function of temperature in comparison with Minkowski spacetime.
Thereby, we are able to comprehend the effects of the curvature caused by the wormhole geometry in the thermodynamic properties of the system. Due to the complexity of the problem, no every value of temperature are allowed. To overcome this situation, hereafter a Wick--like rotation is implemented seeking physical results.

\begin{figure}[tbh]
  \centering
\includegraphics[scale=0.405]{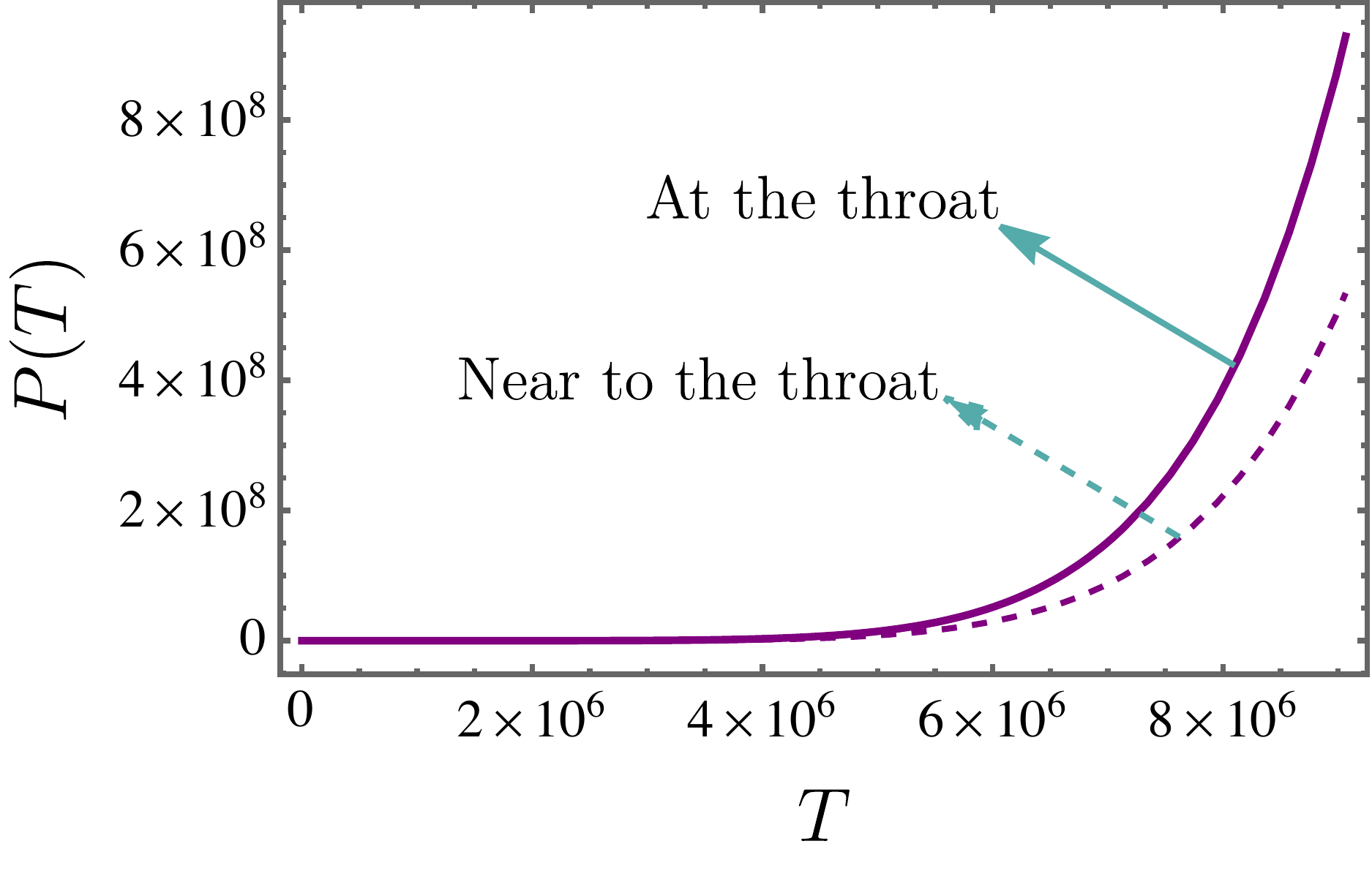}
\includegraphics[scale=0.45]{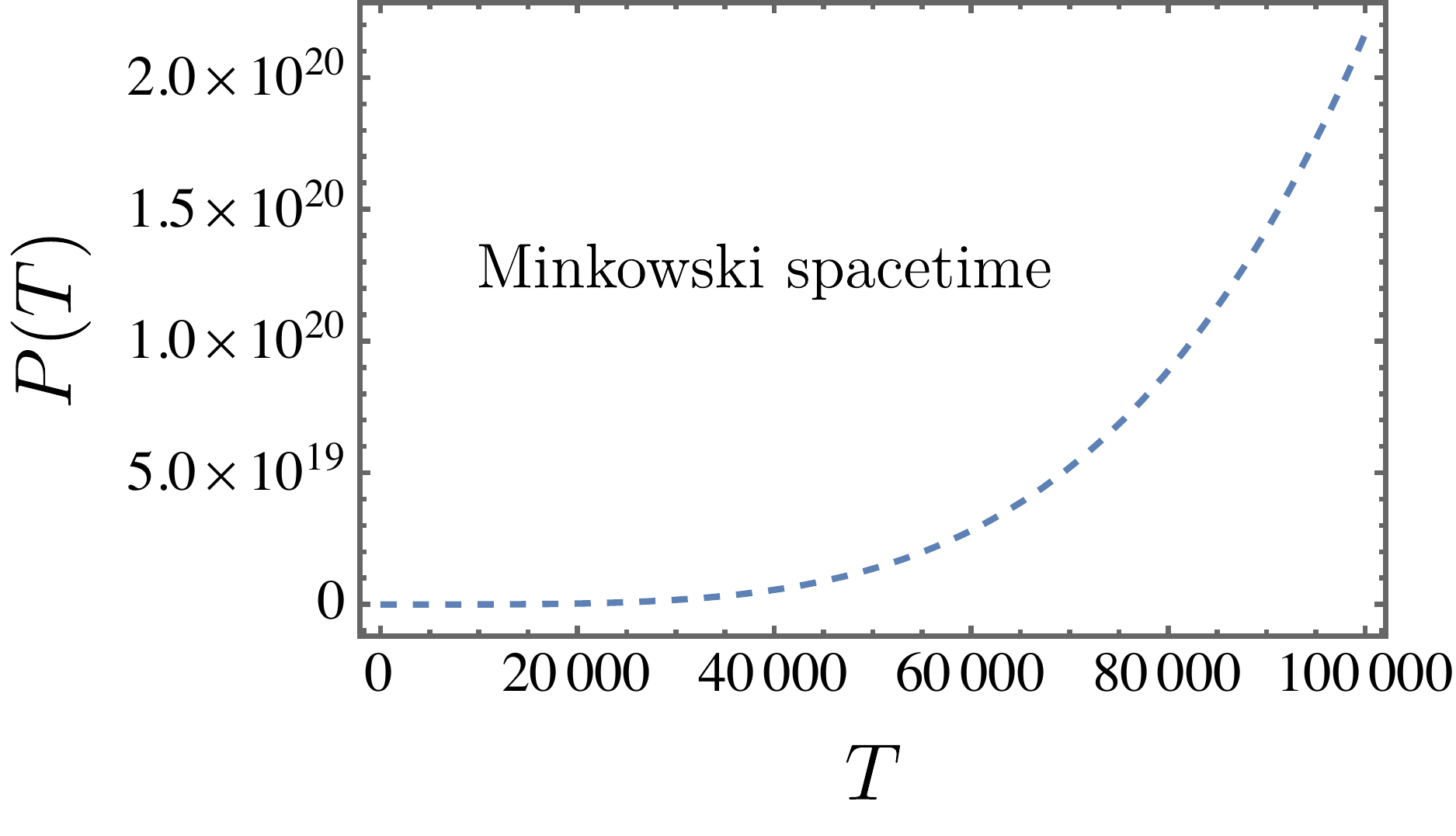}
  \caption{On the left side, we display the pressure as function of temperature, considering two different regions: at the throat, and near to it. Also, the pressure is shown for the Minkowski spacetime on the right hand.}\label{Pressure--comparison}
\end{figure}

Fig. \ref{Pressure--comparison} displays such a comparison. Here, it is important to mention that we separate that plot of the flat spacetime because of scale dimension. Otherwise, the wormhole plots should be just a straight line parallel to the $x$--axis.
Notice that one of the main differences is that, in the usual case, the pressure grows faster than in the wormhole geometry when the temperature rises.

\subsection{Mean energy}

Being an extensive quantity, the mean energy is an indispensable state quantity to address the thermal properties of a given system. As it is well--known, any thermodynamic processes that delineates the mean energy bears with transference of energy, matter, and work. In other words, these features form a sufficient background to deal with the first thermodynamic law. Thereby, we can write such a state function as
\ie
\begin{split}
U(\beta,r,\gamma) = &\int_{0}^{\infty} \frac{E^2}{c_0^3 \hbar  \left(1-e^{-\beta  E}\right)} \left(\frac{B^2}{r^2}+1\right)^3 \\
& \times e^{\left(-\beta  E-8 \gamma  \tan ^{-1}\left(\frac{r}{B}\right)+\frac{1}{2} \left(4 \pi  \gamma -8 \gamma  \tan ^{-1}\left(\frac{r}{B}\right)\right)+4 \pi  \gamma \right)}\\
& \times \sqrt{E^2 \hbar ^2-c_0^4 m_0^2 e^{4 \gamma  \tan ^{-1}\left(\frac{r}{B}\right)-2 \pi  \gamma }}\, \mathrm{d}E.
\end{split}
\label{meanenergyequation}
\fe
Analogously what we have done in the previous sections, here, we shall investigate the consequences of this thermodynamic function to those three different scenarios: asymptotic behavior, at the throat and near to the throat.

\subsubsection{Asymptotic behavior}

Initially, let us consider the asymptotic behavior for the mean energy. In other words, we take into account a situation where $r \rightarrow \infty$. Thereby, with the absence of an analytical solution, such a thermal quantity $U_{a}(\beta,\gamma)  = \lim_{r \to \infty} U(\beta,r,\gamma)$ will be discussed in a numerical manner.

\begin{figure}[tbh]
  \centering
\includegraphics[width=8cm,height=5cm]{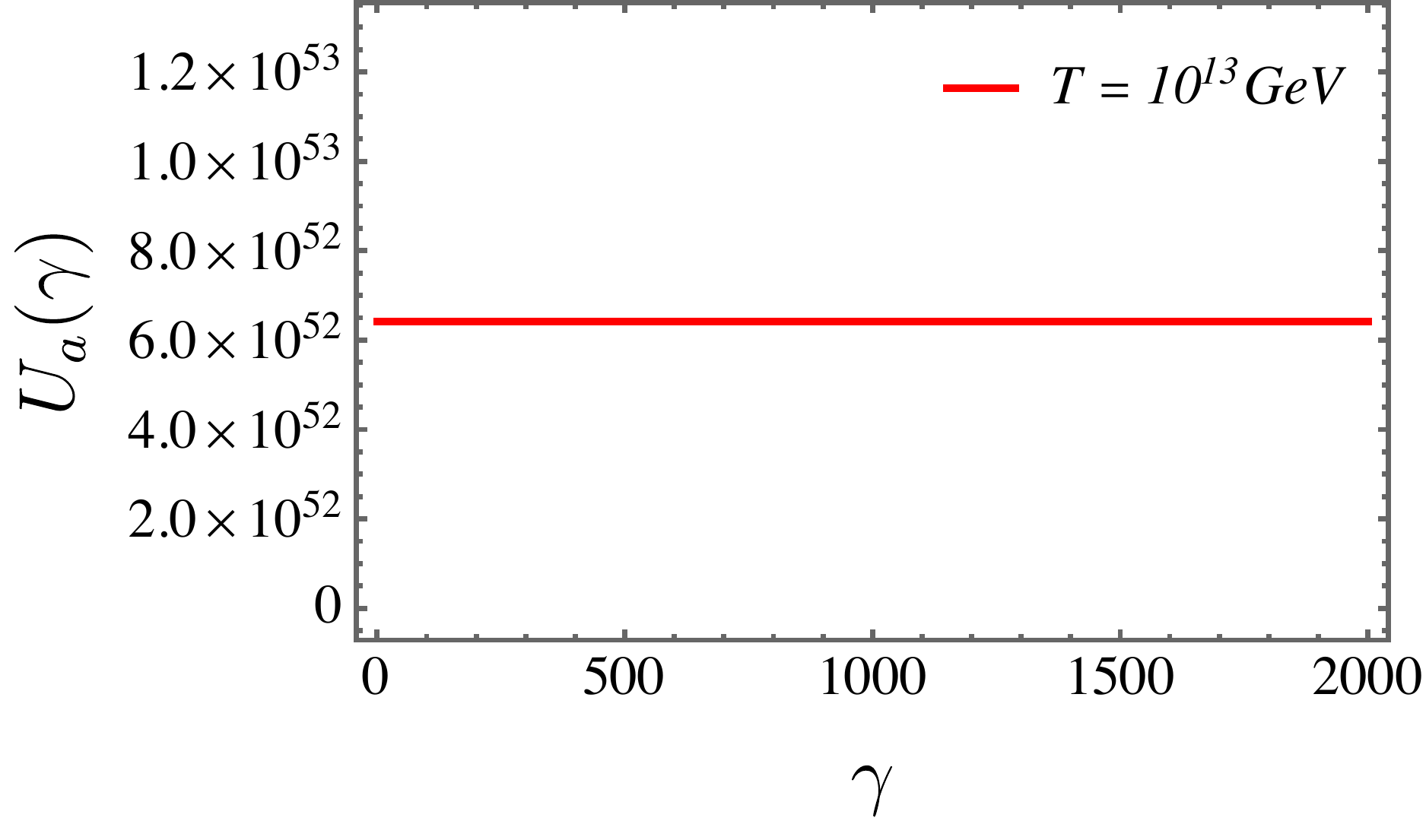}
  \caption{Mean energy considering the asymptotic behavior for the inflationary era ($T=10^{13}$ GeV)}\label{meanenergyequationAsymptoticfig}
\end{figure}

The behavior of $U_{a}(\beta,\gamma)$ is displayed in Fig. \ref{meanenergyequationAsymptoticfig}. We see that there is a constant value of $U_{a}(\beta,\gamma)$ for any value of parameter $\gamma$. Similarly to what happens to the pressure, its constant behavior, to the asymptotic case, is a direct consequence of the Ellis wormhole be asymptotically flat spacetime.

\subsubsection{At the throat}

This subsection is devoted to investigate the consequences of the mean energy at the throat. 

\begin{figure}[tbh]
  \centering
  \includegraphics[width=8cm,height=5cm]{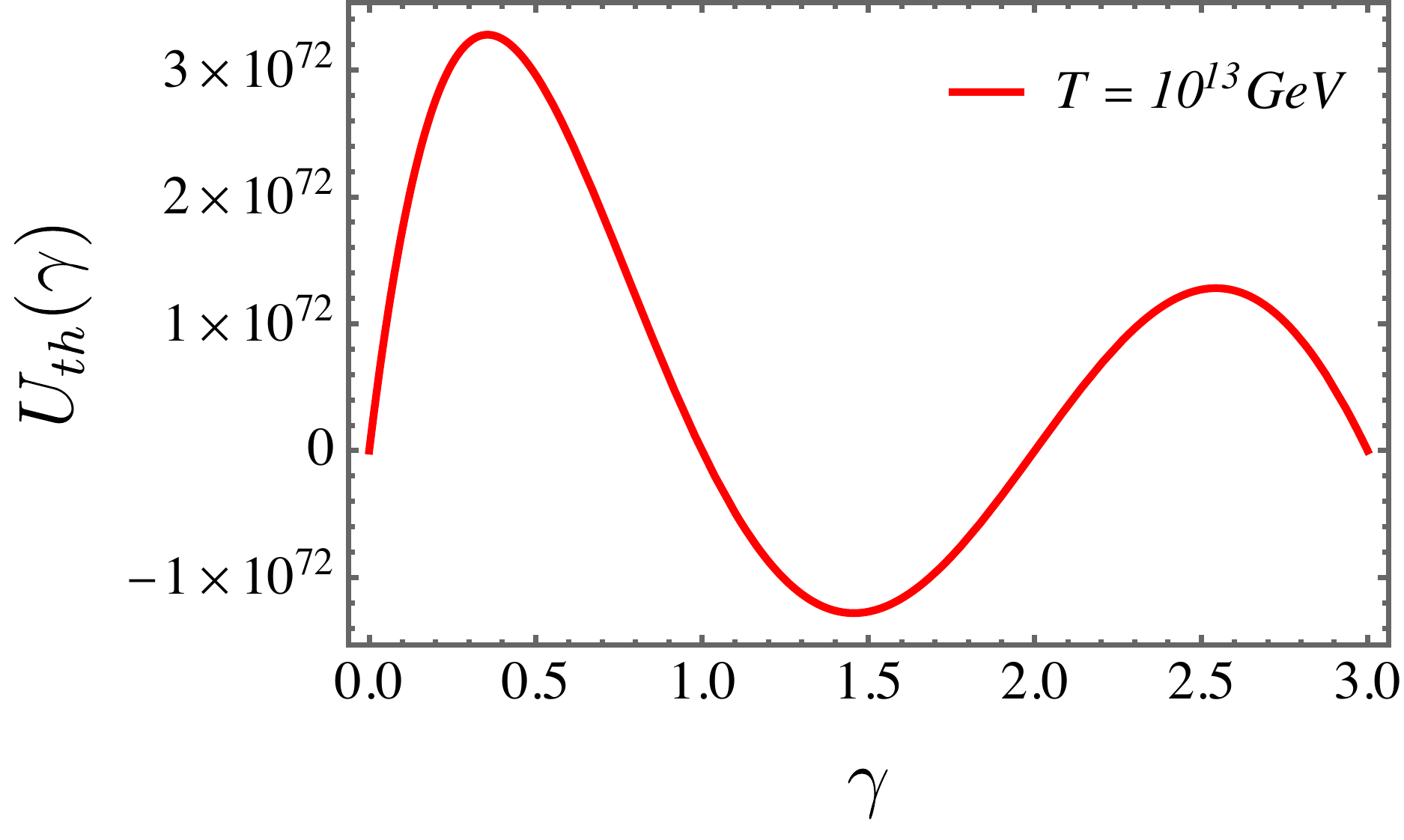}
  \caption{Mean energy regarding a point at the throat for the inflationary era ($T=10^{13}$ GeV)}\label{meanenergyequationthroat}
\end{figure}
The behavior for this configuration is shown in Fig. \ref{meanenergyequationthroat}. Here, we verify that there is a range of values of $\gamma$, namely $1 < \gamma < 2$, in which the mean energy assumes negative values. This clearly indicates an instability of the system caused by the strong effects of the geometry at this point.


\subsubsection{Near to the throat}

\begin{figure*}[ht]
\centering
\begin{subfigure}{.4\textwidth}
  \centering
  \includegraphics[scale=0.4]{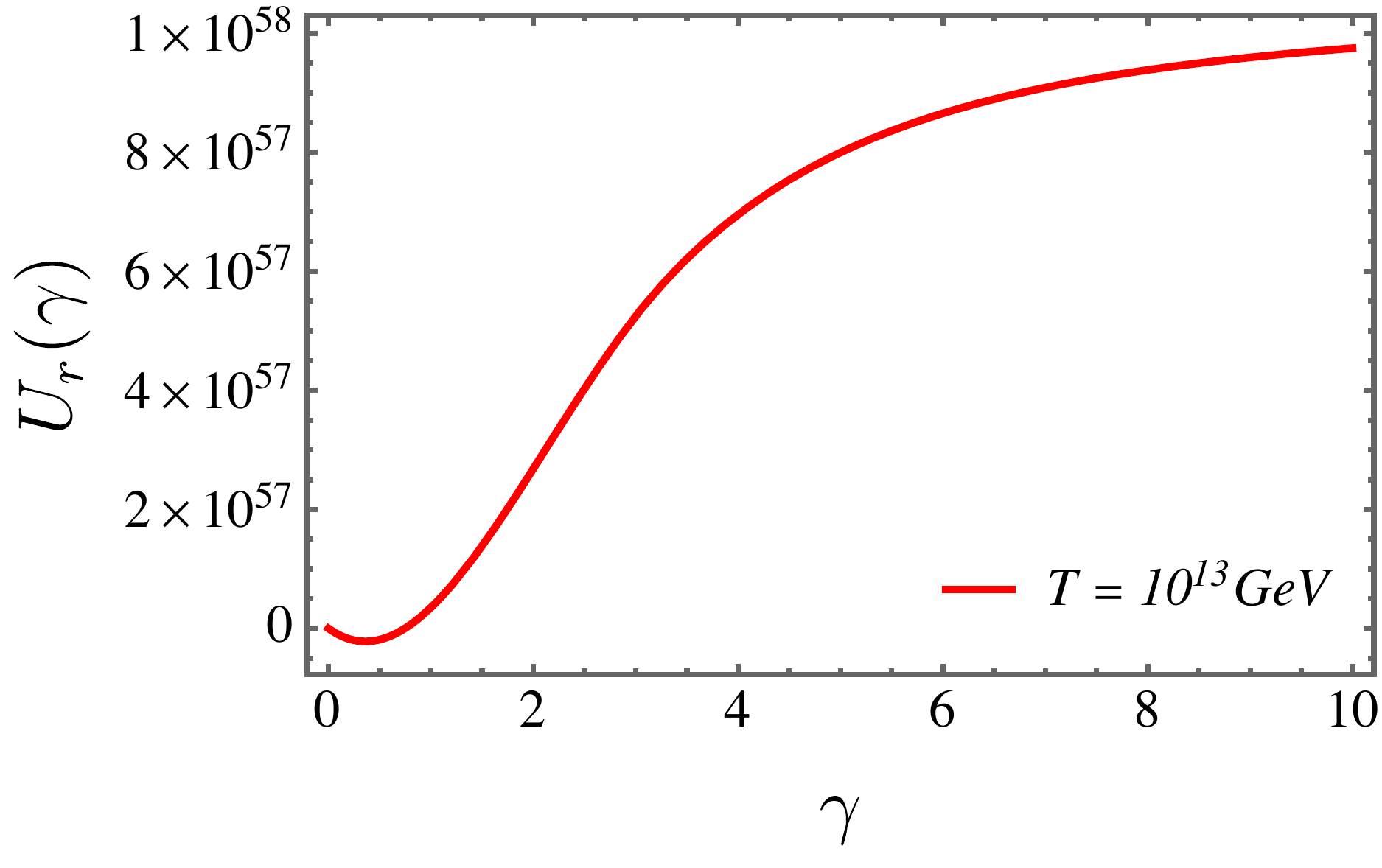}
  \caption{}
\end{subfigure}%
\begin{subfigure}{.6\textwidth}
  \centering
  \includegraphics[scale=0.4]{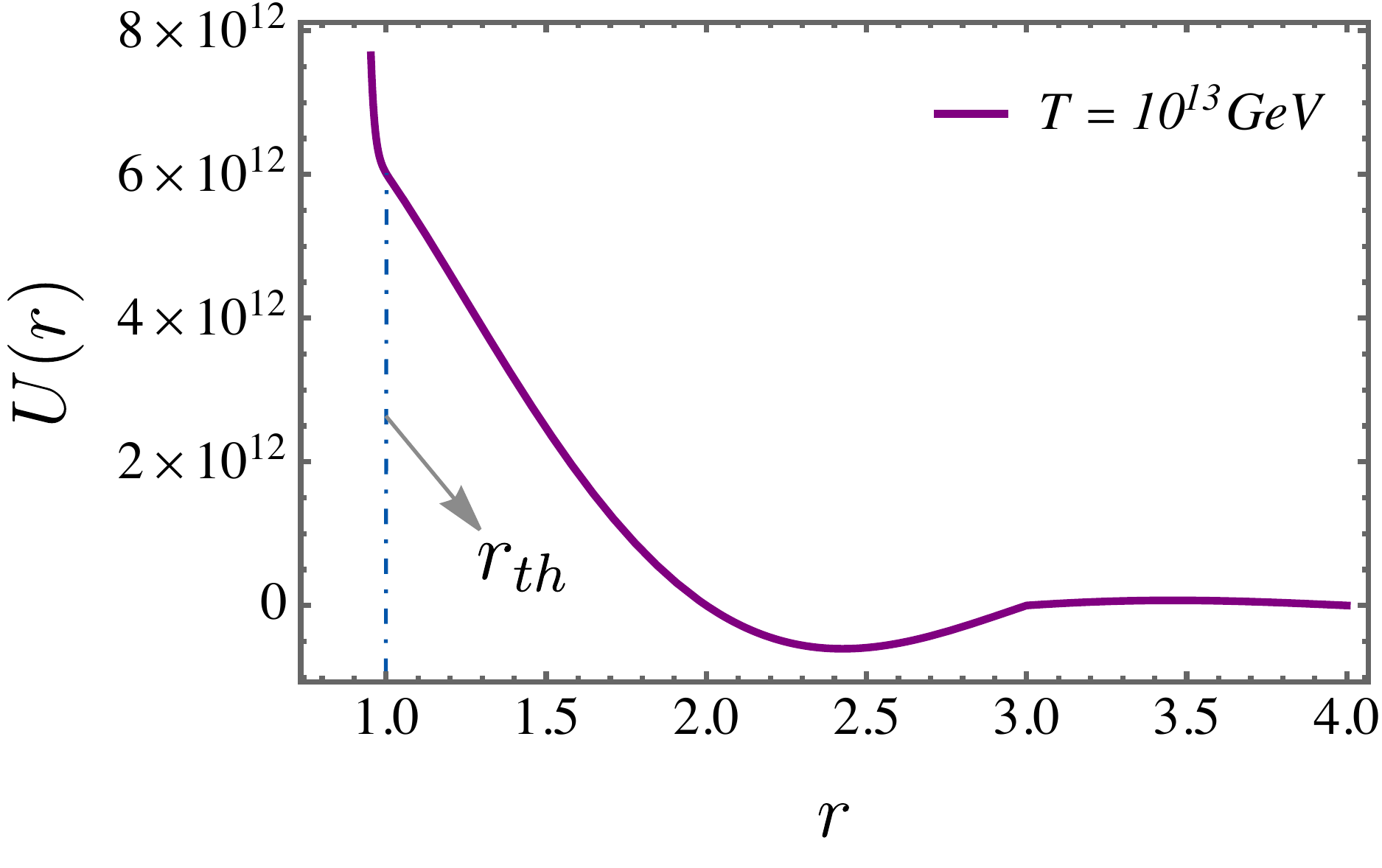}
  \caption{}
 
\end{subfigure}%
\caption{Mean energy regarding a point near to the throat for the inflationary era ($T=10^{13}$ GeV). The panel (a) stands for the mean energy as a function of the parameter $\gamma$, while the panel (b) stands for the mean energy as a function of the radial coordinate $r$.}\label{meanenergyequationnearthroat}
\end{figure*}

This subsection is devoted to explore the main aspects of the mean energy when we consider a point close to the throat, i.e., $r=M\sqrt{1+\gamma^{2}} \gamma$. In Fig. \ref{meanenergyequationnearthroat}, we display how the mean energy behaves to a point close to the throat. In this case, its shape includes a forbidden region in the range $0<\gamma<1$, where $U_{r}(\gamma)$ assumes negative values when we consider the inflationary regime of temperature of the universe.

On the right hand of Fig. \ref{meanenergyequationnearthroat}, we show the mean energy, $U(r)$, a function of the radial coordinate $r$ only. Corroborating our previous results to the pressure, the mean energy similarly possesses negative values for that particular choice of parameters $\gamma$, i.e., $2<\gamma<3$.

\begin{figure}[tbh]
  \centering
\includegraphics[scale=0.405]{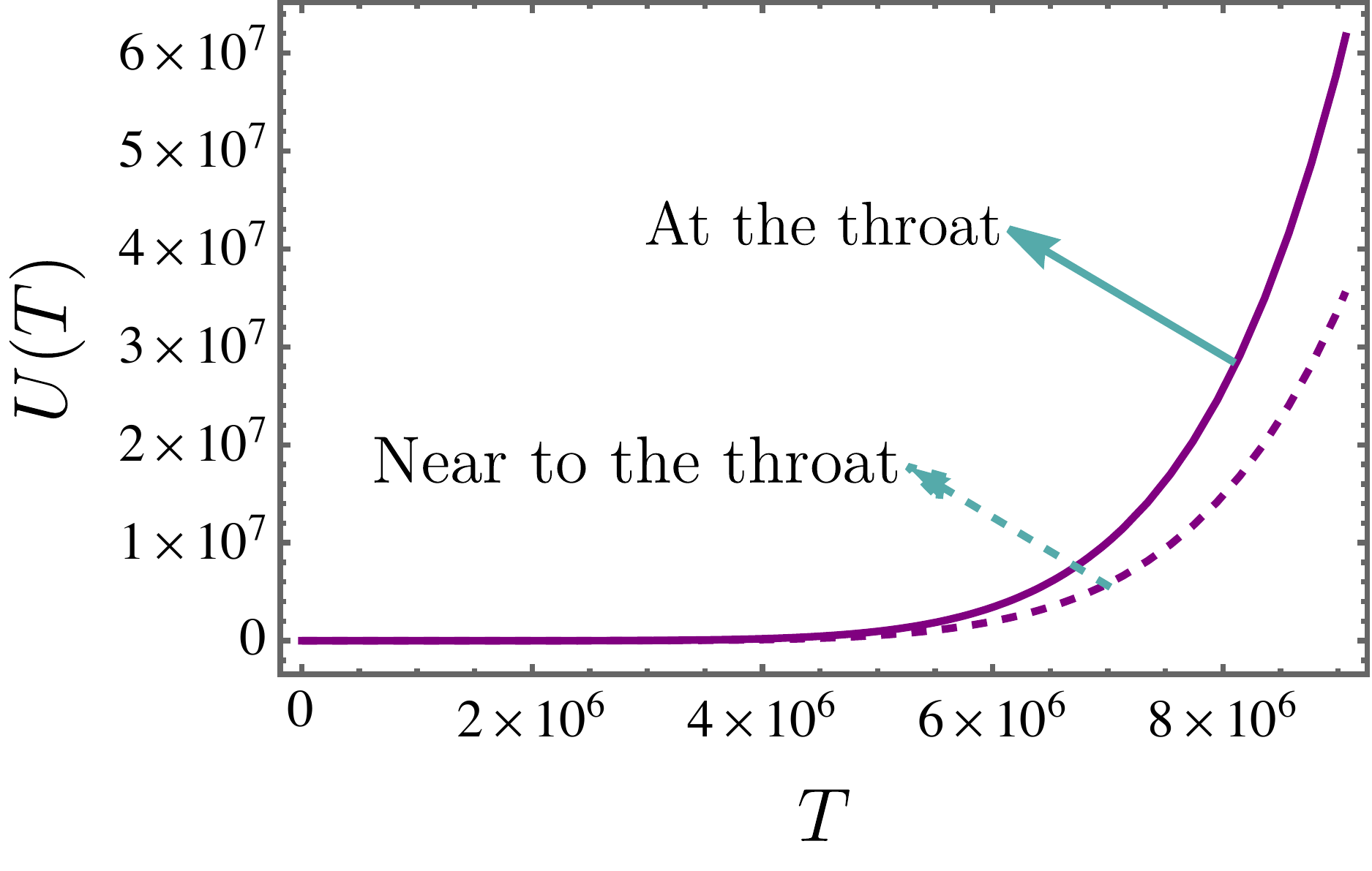}
\includegraphics[scale=0.45]{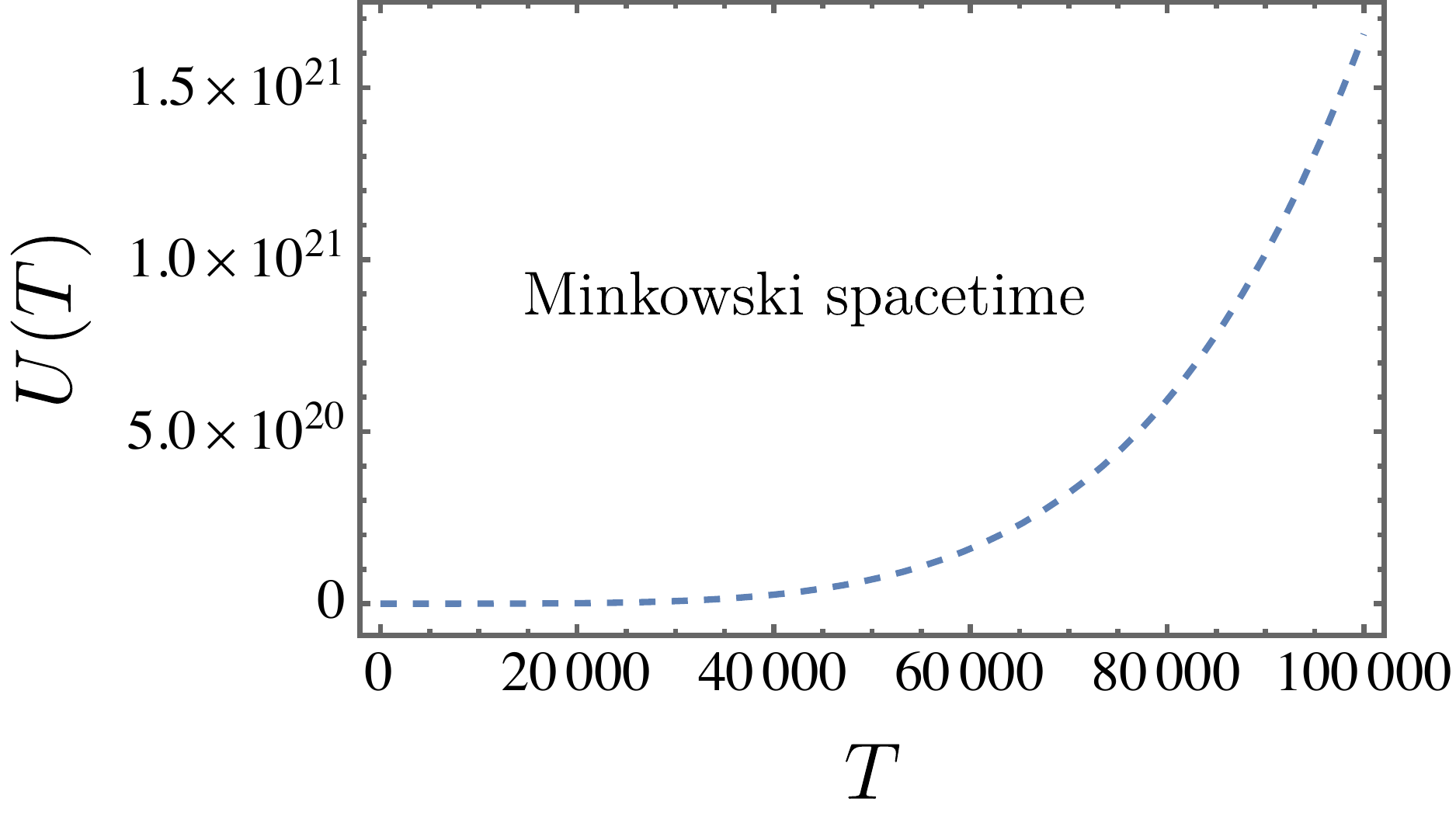}
  \caption{On the left side, we display the mean energy as function of temperature, considering two different regions: at the throat, and near to it. Such a thermal quantity is also shown for the Minkowski spacetime on the right hand.}\label{energy--comparison}
\end{figure}

Additionally, another dimension that merits investigation is the examination of how mean energy varies as a function of temperature, especially in contrast to our earlier observations in Minkowski spacetime. This enables us to better understand the thermodynamic implications of the curvature introduced by the wormhole geometry.

Figure \ref{energy--comparison} illustrates this comparison. It is worth noting that we have separated the plot for flat spacetime to accommodate differences in scale dimensions; otherwise, the wormhole curves would appear as a straight line parallel to the $x$--axis. A key distinction to observe is that, in the Minkowski case, the rate at which mean energy increases with temperature is faster than it is in the case of wormhole geometry.


\subsection{Entropy}

Whenever we want to know about the reversibility of a given process, one reasonable approach is considering the thermodynamic state quantity called entropy. With it, we can infer about many thermal process of field theories. Such a function brings out the knowledge of the second law of thermodynamics. In this way, the entropy can be written as 
\begin{widetext}
\ie
\begin{split}
S(\beta,r,\gamma) =& \int_{0}^{\infty}  \frac{\beta}{  c_0^3 \hbar  \left(1-e^{-\beta  E}\right)} E^2 \left(\frac{B^2}{r^2}+1\right)^3 e^{-\beta  E-8 \gamma  \tan ^{-1}\left(\frac{r}{B}\right)+\frac{1}{2} \left(4 \pi  \gamma -8 \gamma  \tan ^{-1}\left(\frac{r}{B}\right)\right)+4 \pi  \gamma} \\
& \times \sqrt{E^2 \hbar ^2-c_0^4 m_0^2 e^{4 \gamma  \tan ^{-1}\left(\frac{r}{B}\right)-2 \pi  \gamma }} \mathrm{d}E - \int^{\infty}_{0}\frac{1}{ c_0^3 \hbar}E \left(\frac{B^2}{r^2}+1\right)^3 \ln \left(1-e^{-\beta  E}\right)\\
& \times e^{-8 \gamma  \tan ^{-1}\left(\frac{r}{B}\right)+\frac{1}{2} \left(4 \pi  \gamma -8 \gamma  \tan ^{-1}\left(\frac{r}{B}\right)\right)+4 \pi  \gamma } \sqrt{E^2 \hbar ^2-c_0^4 m_0^2 \,e^{4 \gamma  \tan ^{-1}\left(\frac{r}{B}\right)-2 \pi  \gamma }} \,\mathrm{d}E.
\end{split}\fe
\end{widetext}
Below, as we have done throughout the manuscript, we shall provide a similar analysis accomplished to the other thermodynamic functions.


\subsubsection{Asymptotic behavior}

\begin{figure}[tbh]
  \centering  \includegraphics[width=8cm,height=5cm]{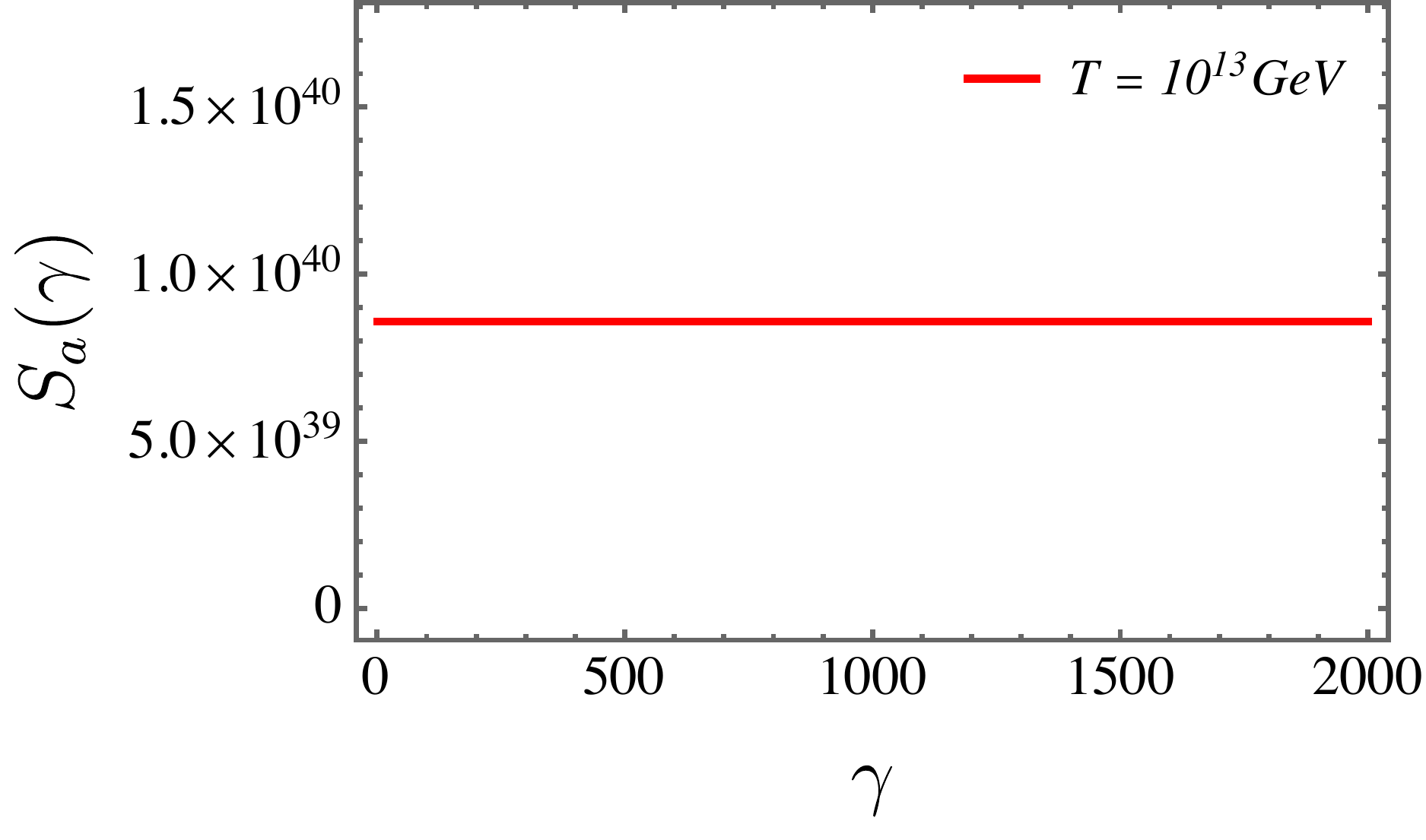}
  \caption{Entropy for the asymptotic behavior considering the inflationary era ($T=10^{13}$ GeV)}\label{entropyAsymptoticfig}
\end{figure}

In this subsection, we present the analysis to the entropy within the asymptotic configuration. In this manner, we take into account the limit where $S_{a}(\beta,\gamma)  = \lim_{r \to \infty} S(\beta,r,\gamma)$. In Fig. \ref{entropyAsymptoticfig}, $S_{a}(\gamma)$ is plotted as a function of parameter $\gamma$, which exhibits constant values of $S_{a}(\gamma)$. The reason for that lies in the same fact that we have previously argued, i.e., the flatness aspect of the spacetime far from the wormhole throat.

\subsubsection{At the throat}

\begin{figure}[tbh]
  \centering
 \includegraphics[width=8cm,height=5cm]{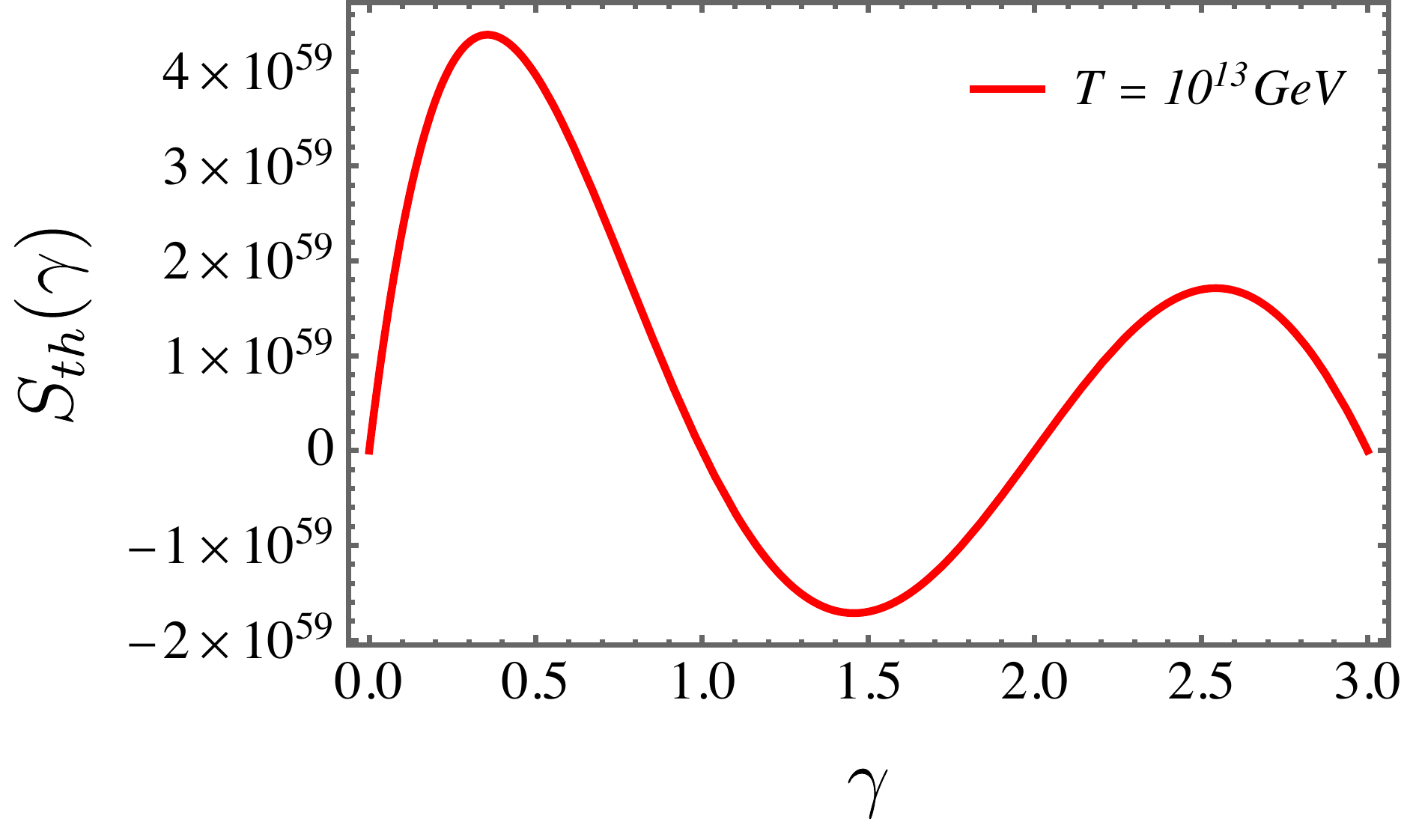}
\caption{Entropy for the point localized at the throat of the wormhole considering the inflationary era ($T=10^{13}$ GeV)}\label{entropythroatfig}
\end{figure}

Here, we obtain the results concerning the entropy to a point localized at the throat of the wormhole, which are displayed in Fig. \ref{entropythroatfig}.
To this configuration, analogously with what happens to the mean energy, when $\gamma$ parameter lies between $1<\gamma<2$, the entropy assumes negative values, indicating instabilities as well. Such a behavior highlights that, at the throat, this range is not allowed for our system.


\subsubsection{Near to the throat}

\begin{figure*}[ht]
\centering
\begin{subfigure}{.4\textwidth}
  \centering
  \includegraphics[scale=0.4]{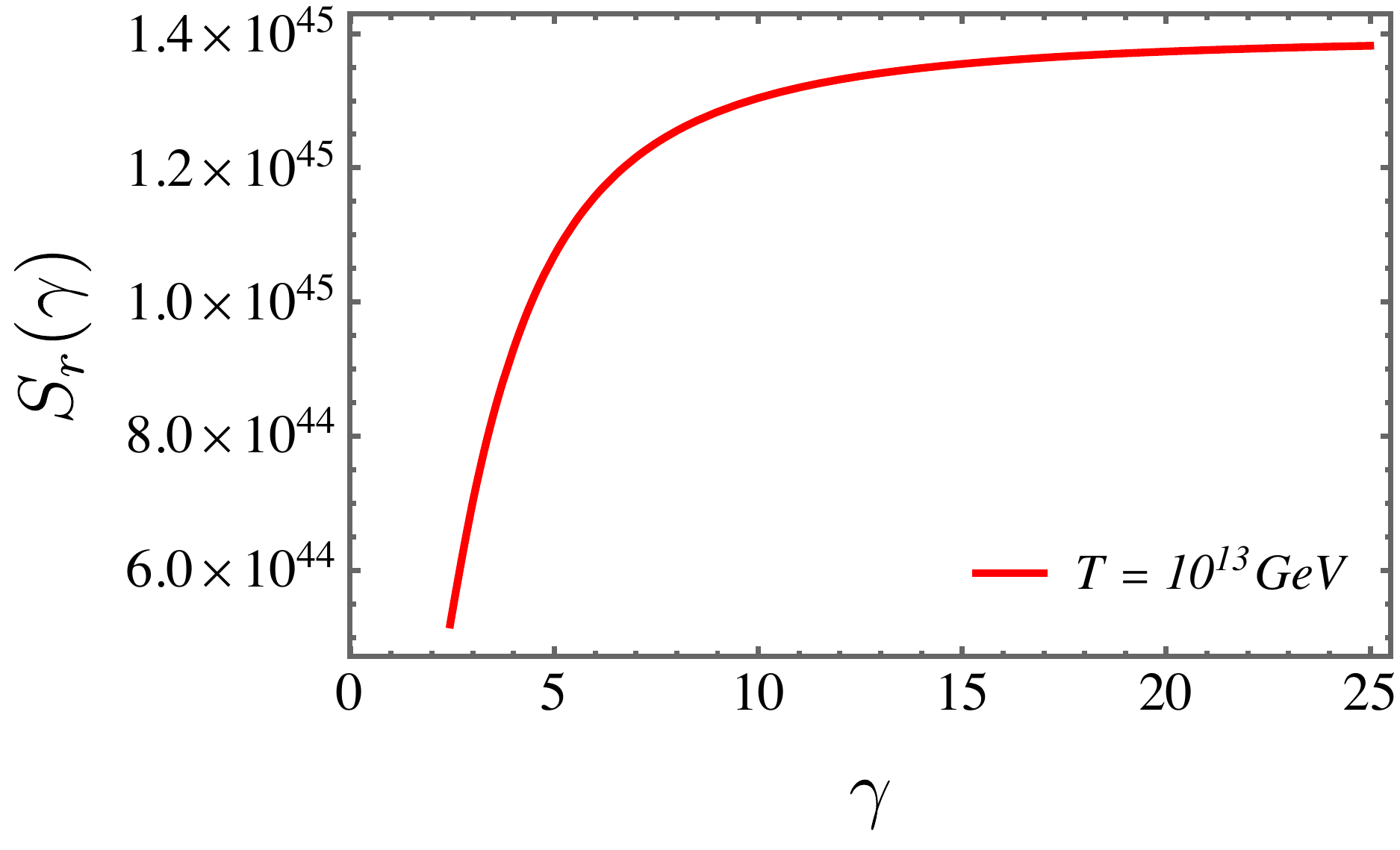}
  \caption{}
\end{subfigure}%
\begin{subfigure}{.6\textwidth}
  \centering
  \includegraphics[scale=0.4]{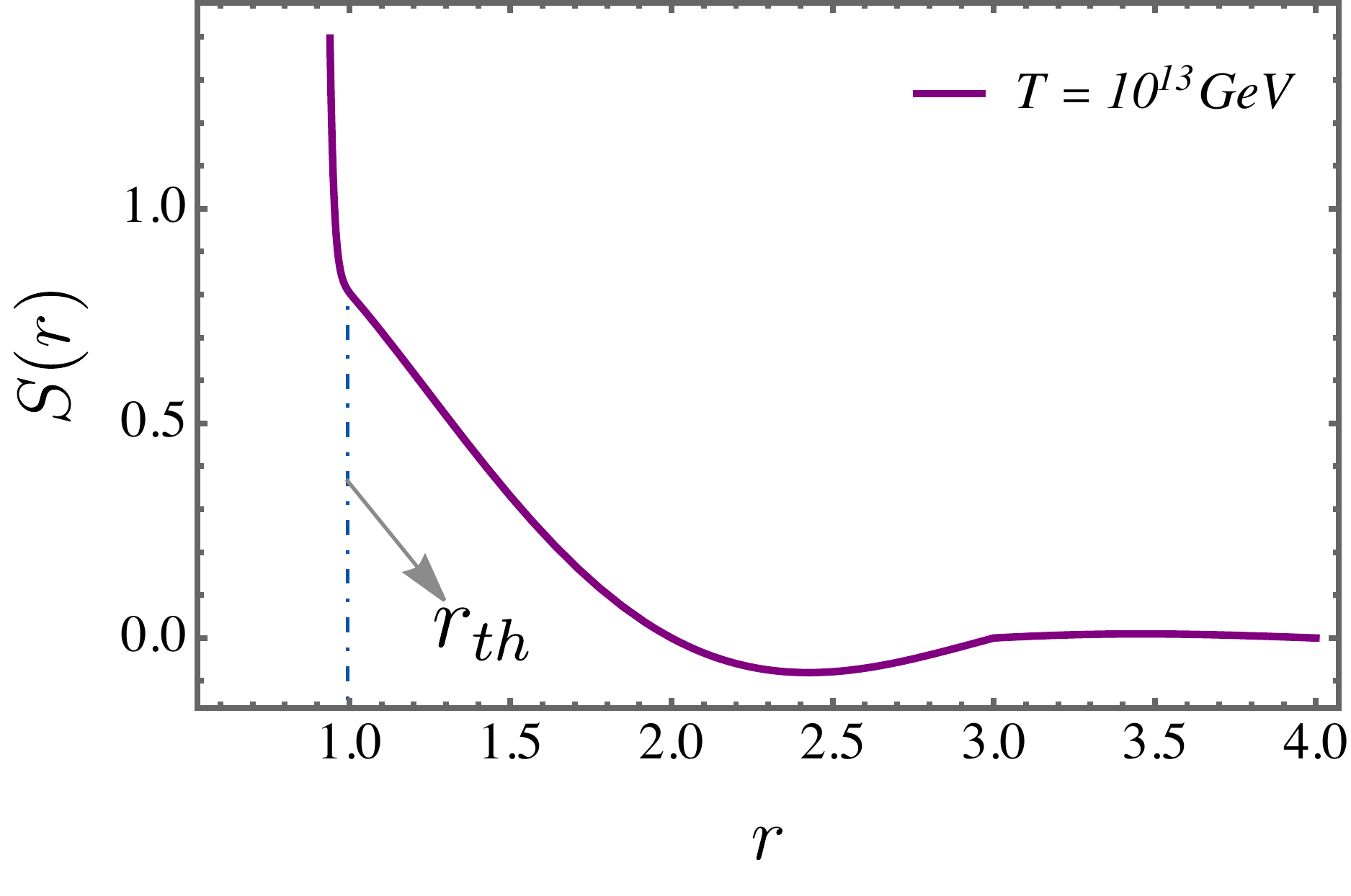}
  \caption{}
 
\end{subfigure}%
\caption{Entropy for the point localized near the throat of the wormhole considering the inflationary era ($T=10^{13}$ GeV). The panel (a) stands for the entropy as a function of parameter $\gamma$, while the panel (b) represents it as a function of the radial coordinate $r$.}\label{entropynearthroatfig}
\end{figure*}

In order to verify the modifications of the entropy to a point close to the wormhole throat, we provide this subsection. The plots depicted in Fig. \ref{entropynearthroatfig} are concerning the inflationary era regime of temperature. Here, we present $S_{r}(\gamma)$, which represents the entropy for a region nearby the throat, and $S(r)$ for the same thermodynamic quantity as a function of the radius coordinate $r$. Notice that, for $2<r<3$, we have negative values for the entropy, indicating, then, the appearance of instabilities. In other words, we have a ``forbidden'' range to the evaluation of the respective thermal properties our system composed fundamentally by scalar particles. 

\begin{figure}[tbh]
  \centering
\includegraphics[scale=0.405]{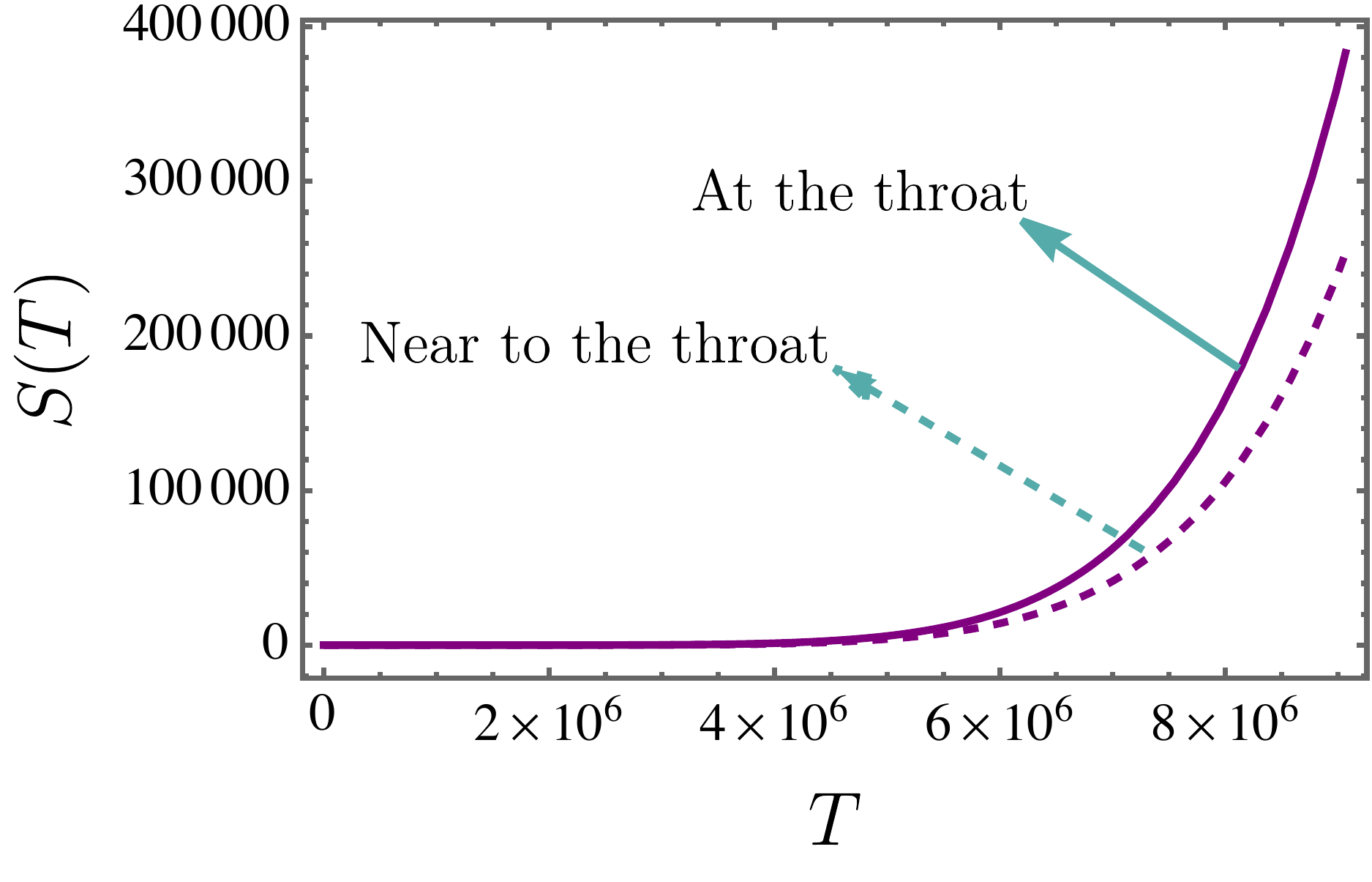}
\includegraphics[scale=0.42]{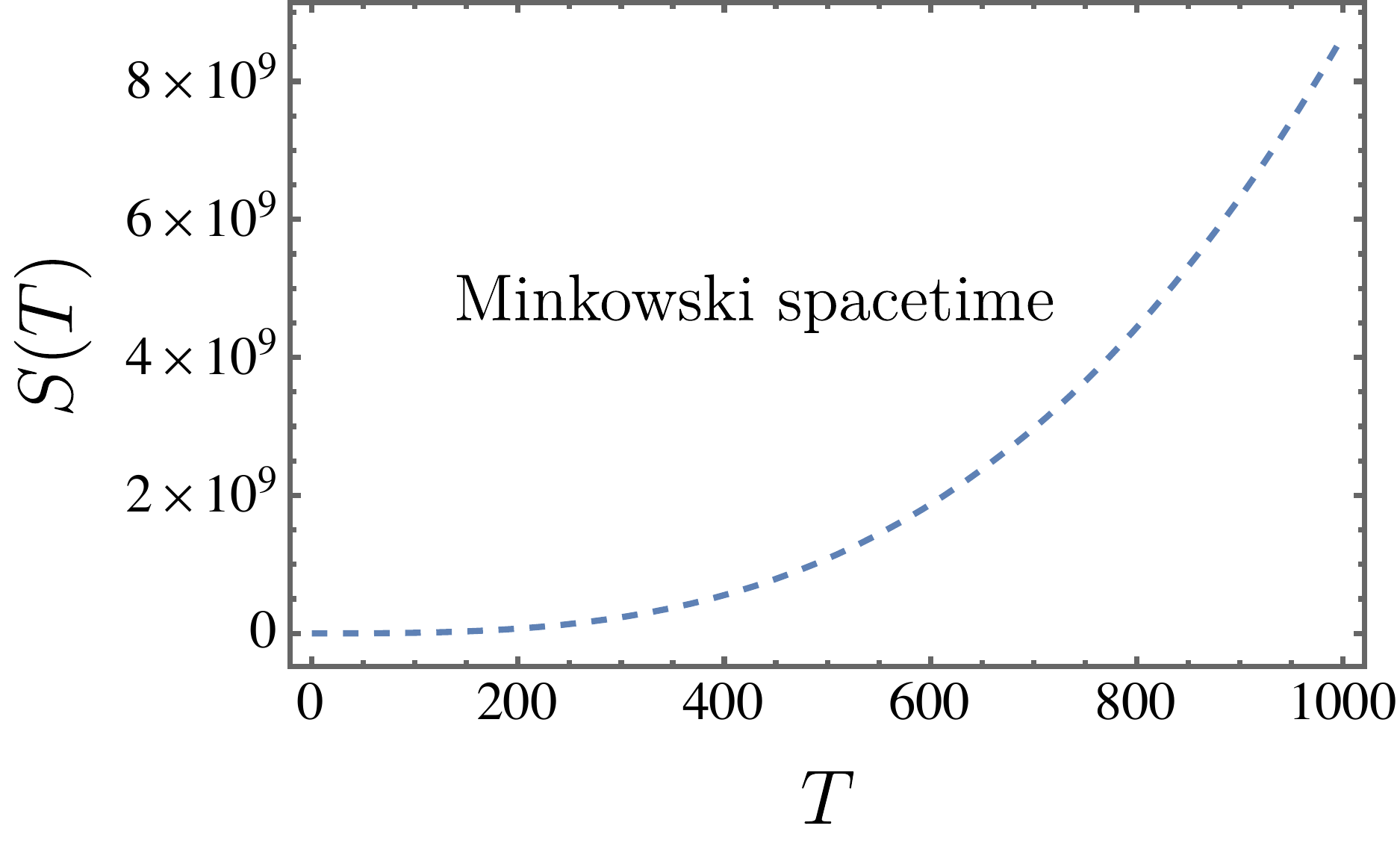}
  \caption{On the left--hand side, the graph illustrates how entropy varies with temperature in two distinct regions: at the wormhole's throat and in its vicinity. For comparison, the thermal properties corresponding to Minkowski spacetime are presented on the right--hand side.}\label{entropy--comparison}
\end{figure}

Another aspect that warrants further study is the relationship between entropy and temperature, especially when juxtaposed with our previous findings in Minkowski spacetime. Exploring this relationship provides a deeper understanding of the thermodynamic effects of the curvature specific to wormhole geometry.

As depicted in Figure \ref{entropy--comparison}, we have chosen to isolate the plot for flat spacetime due to differences in scale dimensions. Otherwise, the curves representing wormhole geometry would essentially appear as a straight line parallel to the \(x\)--axis. An important observation is that, in the context of Minkowski spacetime, the rate at which entropy increases with temperature is faster than in the case of wormhole geometry.


\subsection{Heat capacity}

Finally, to finish our discussion involving the thermodynamic properties of the Ellis wormhole, we present the reaming thermodynamic function, the heat capacity. In this sense, we write
\begin{widetext}
\ie
\begin{split}
C(\beta,r,\gamma) = & \int^{\infty}_{0}
\frac{\beta^{2}E^3}{c_0^3 \hbar  \left(1-e^{-\beta  E}\right)}\left(\frac{B^2}{r^2}+1\right)^3 \sqrt{E^2 \hbar ^2-c_0^4 m_0^2 e^{4 \gamma  \tan ^{-1}\left(\frac{r}{B}\right)+2 \epsilon }} \\
 &\times  e^{\left(-\beta  E+\frac{1}{2} \left(-8 \gamma  \tan ^{-1}\left(\frac{r}{B}\right)+2 \zeta -2 \epsilon \right)-8 \gamma  \tan ^{-1}\left(\frac{r}{B}\right)+2 \zeta -2 \epsilon \right)} \mathrm{d}E \\
 & + \int^{\infty}_{0} \frac{\beta^{2} E^3}{c_0^3 \hbar  \left(1-e^{-\beta  E}\right)^2} \left(\frac{B^2}{r^2}+1\right)^3 \sqrt{E^2 \hbar ^2-c_0^4 m_0^2 e^{4 \gamma  \tan ^{-1}\left(\frac{r}{B}\right)+2 \epsilon }} \\
 & \times  e^{\left(-2 \beta  E+\frac{1}{2} \left(-8 \gamma  \tan ^{-1}\left(\frac{r}{B}\right)+2 \zeta -2 \epsilon \right)-8 \gamma  \tan ^{-1}\left(\frac{r}{B}\right)+2 \zeta -2 \epsilon \right)}\mathrm{d}E
\end{split}
\fe
\end{widetext}

\subsubsection{Asymptotic behavior}

\begin{figure}[tbh]
  \centering
\includegraphics[width=8cm,height=5cm]{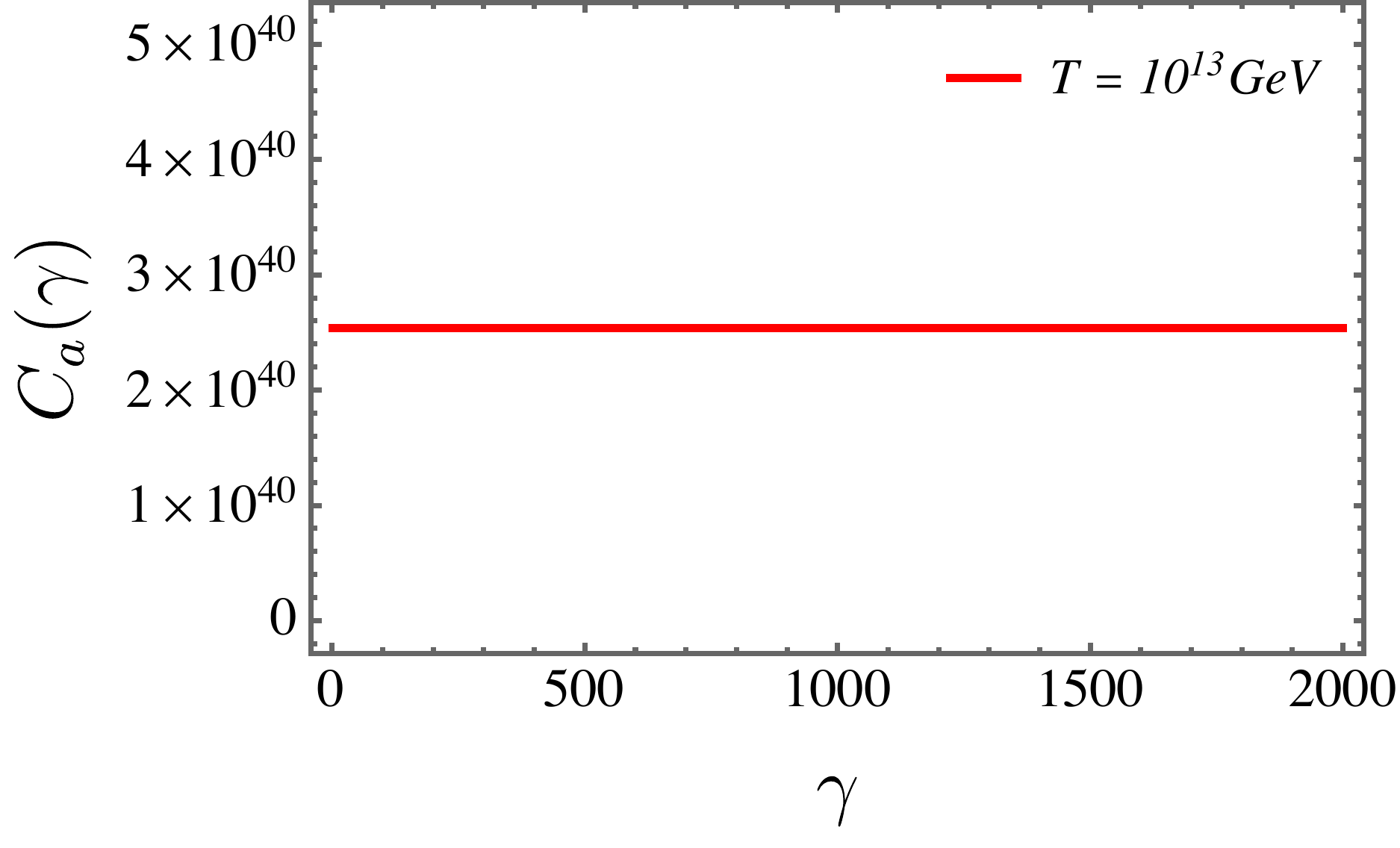}
  \caption{The heat capacity for the point localized asymptotically in comparison to the wormhole considering the inflationary era ($T=10^{13}$ GeV)}\label{heatcapacityaAsymptoticbehavior}
\end{figure}

Here, we provide the study to the heat capacity for the asymptotic case. Thereby, we label $C_{a}(\beta,\gamma) = \lim_{r \to \infty} C(\beta,r,\gamma)$. In this case, we have also a constant behavior for it, which uncovers the existence of a flatness aspect to the wormhole spacetime, i.e., far from the throat. This facet is shown in Fig. \ref{heatcapacityaAsymptoticbehavior}. 

\subsubsection{At the throat}

The heat capacity at the throat shows a region possessing negative values as well, when the range $\gamma$ between $1<\gamma<2$ is taken into account. As already mentioned, this indicates the appearance of ``forbidden'' values to our particles system.

\begin{figure}[tbh]
  \centering
 \includegraphics[width=8cm,height=5cm]{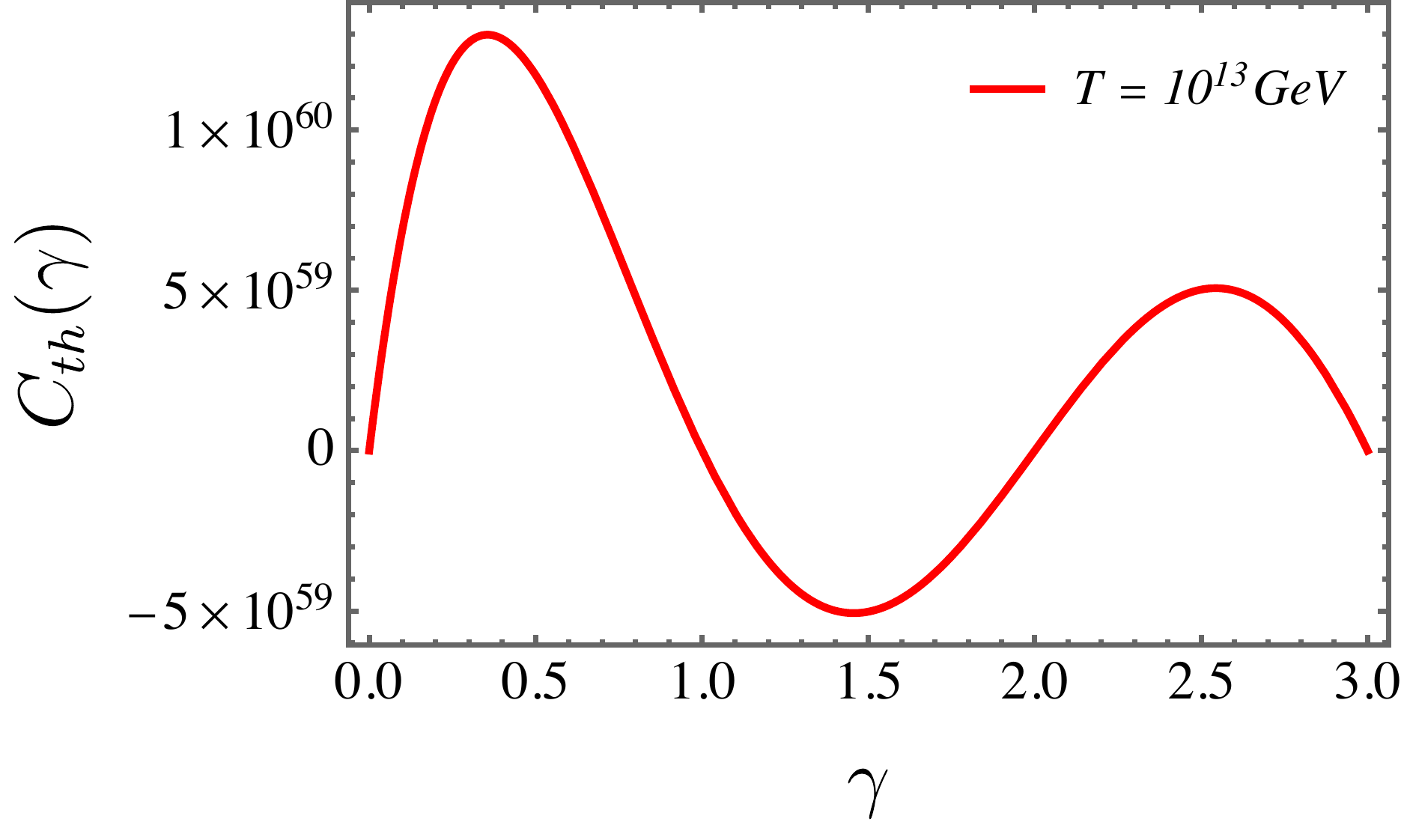}
  \caption{The heat capacity for the point localized at the throat of the wormhole considering the inflationary era ($T=10^{13}$ GeV)}\label{heatcapacityathroat}
\end{figure}


\subsubsection{Near to the throat}

\begin{figure*}[ht]
\centering
\begin{subfigure}{.4\textwidth}
  \centering
  \includegraphics[scale=0.4]{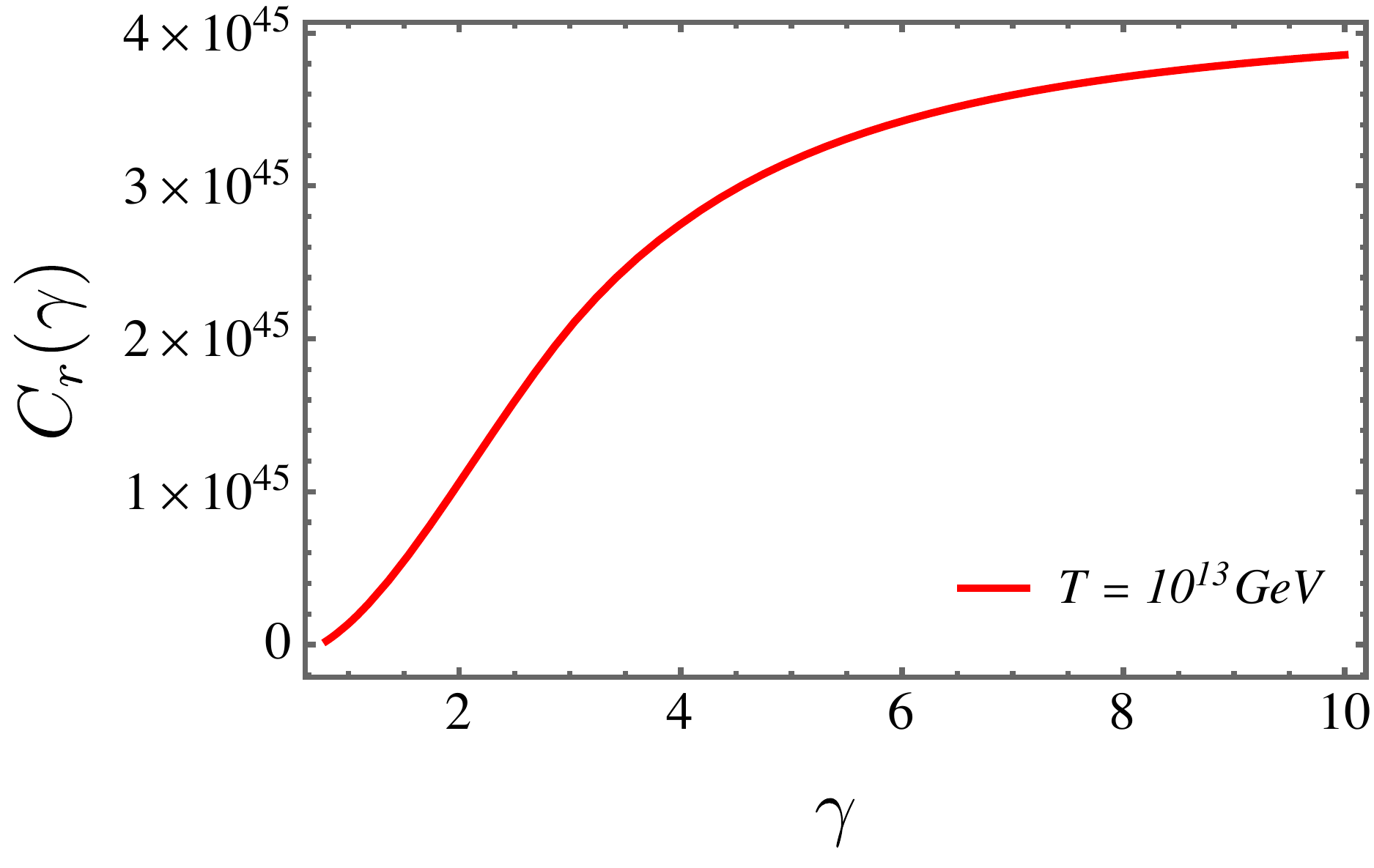}
  \caption{}
\end{subfigure}%
\begin{subfigure}{.6\textwidth}
  \centering
  \includegraphics[scale=0.4]{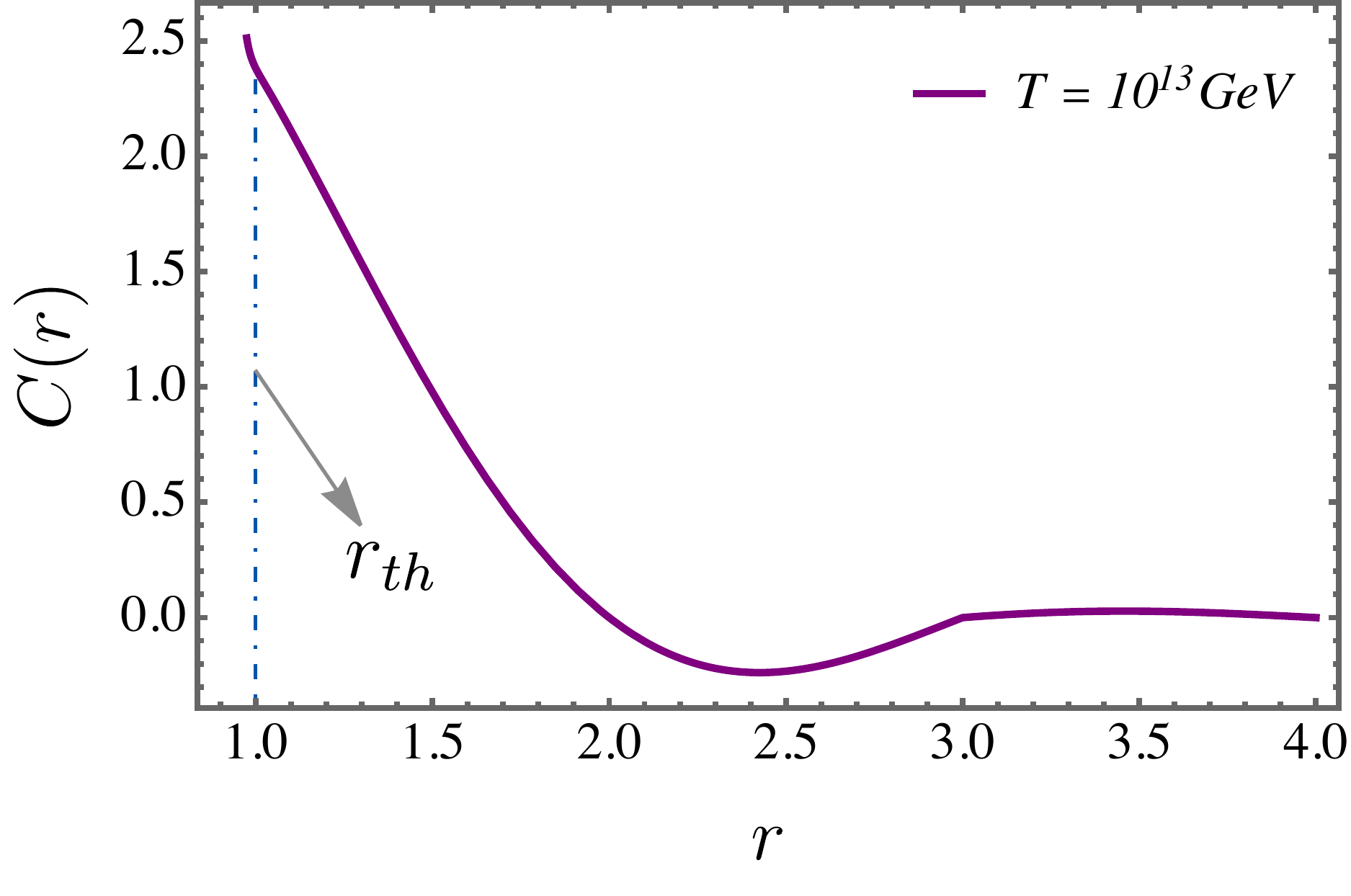}
  \caption{}
 
\end{subfigure}%
\caption{Heat capacity for the point localized near the throat of the wormhole considering the inflationary era ($T=10^{13}$ GeV). The panel (a) stands for the entropy as a function of parameter $\gamma$, while the panel (b) regards the entropy as a function of the radial coordinate $r$.}\label{heatcapacityaneartothethroat}
\end{figure*}

Now, we examine the consequences to the heat capacity for the a point displayed near to the throat. Similarly to the other thermodynamic quantities, the heat capacity also exhibit negative values. The plots depicted in Fig. \ref{heatcapacityaneartothethroat} rile on inflationary era regime of temperature. As we can see, the heat capacity becomes negative when $2<r<3$. This may possibly indicate a phase transition between these values.

\begin{figure}[tbh]
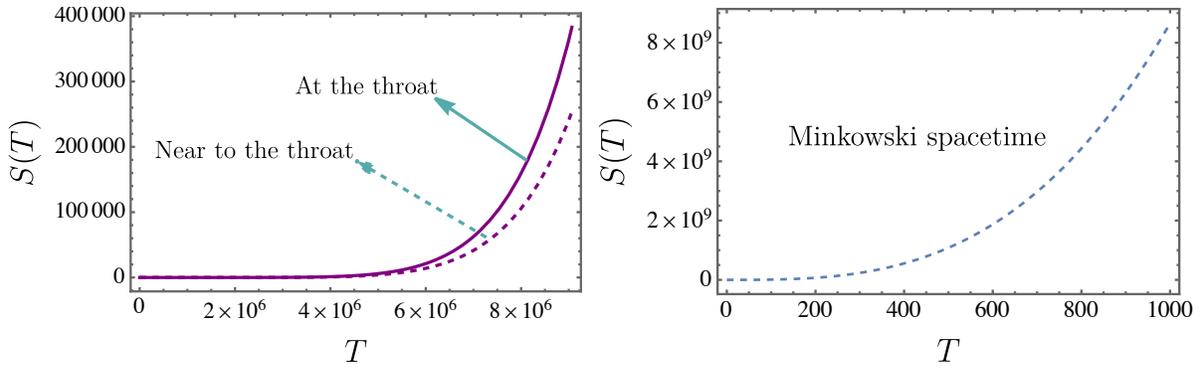

  \centering
\includegraphics[scale=0.405]{entropy-comparison.pdf}
\includegraphics[scale=0.42]{entropy-usual.pdf}
  \caption{On the left-hand side of the graph, we display the variation of heat capacity as a function of temperature, focusing on two specific regions: at the wormhole's throat and near it. To provide a point of comparison, the graph on the right-hand side shows the corresponding thermal properties in Minkowski spacetime.}\label{heatcapacity--comparison}
\end{figure}

Finally, we focus on how the heat capacity is related to the temperature, especially when contrasted with our earlier results in Minkowski spacetime. Such exploration would offer a more nuanced understanding of the thermodynamic consequences tied to the unique curvature of wormhole geometry.

As shown in Figure \ref{heatcapacity--comparison}, we opted to separate the plot for Minkowski spacetime due to varying scale dimensions. If not for this separation, the curves for wormhole geometry would appear almost as a straight line parallel to the \(x\)-axis. A noteworthy point is that in Minkowski spacetime, heat capacity escalates with temperature at a quicker rate compared to its behavior in wormhole geometry.


\section{Interactions} \label{Interaction}

This section focuses on the examination of particle interactions. To achieve this, we enhance the existing methodology by incorporating an interaction term denoted as $U\left( V,n\right)$. Importantly, this term is solely reliant on the values of volume $V$ and particle density $n$ for our specific objectives. As we delve into this topic, we will observe that this interaction can be approximated through the commonly used mean field approximation. This approach facilitates the derivation of analytical outcomes. Notably, it's worth highlighting that the interaction term varies monotonically based on particle density. We make the assumption that $Vu\left( n\right)$ exhibits linearity concerning $\sum_{\Omega}N_{\Omega}=N$, which aids in simplifying the computations. The Taylor expansion of $u\left( n\right)$ (around the mean value $\bar{n}$) is performed as follows:

\begin{equation}
u\left( n\right) =u\left( \bar{n}\right) +u^{\prime }\left( \bar{n}\right)
\left( n-\bar{n}\right) +\ldots. \label{eq:Taylor_u}
\end{equation}%
Furthermore, a crucial observation to emphasize is that when the potential energy solely relies on the position, the \textit{molecular field approximation} becomes a fitting method to employ starting from Eq. (\ref{eq:Taylor_u}). This technique finds extensive utilization in the realm of condensed matter physics, particularly within the existing literature.\cite{humphries1972,klein1969,das2016particle,wojtowicz,ter1962molecular,araujo2017,silva2018,araujo2022does,reis2021thermal,ghosh1,ghosh2}.

Drawing from the comprehensive exploration of this technique in the reference \cite{reis2021thermal}, we are empowered to introduce an interaction term into our scenario. This incorporation gives rise to what is known as the grand canonical potential, accompanied by its interaction component, denoted as follows:
\begin{eqnarray}
\Phi  &=&-T\ln \mathcal{Z} \notag\\
&=& T  \int_{0}^{\infty} {\bf{k}}^{2}\,\mathrm{d}{\bf{k}}\phantom{a}\ln \left( 1+ -\exp \left[ -\beta \left( E+u^{\prime }\left( \bar{n}\right) -\mu \right) \right]\right) +U\left( V,\bar{n}\right) -u^{\prime }\left( \bar{n}\right) \bar{N}.
\label{eq:GCP-new}
\end{eqnarray}

From above equation, we initiate our exploration by considering the mean particle number, which is expressed as follows:

\begin{equation}
\bar{N}= \int_{0}^{\infty} {\bf{k}}^{2}\,\mathrm{d}{\bf{k}}\phantom{a}\frac{1}{\exp \left[ \beta \left( E+u^{\prime }\left( \bar{n}\right) -\mu \right) \right] -1 }.
\label{eq:Mean-number-N}
\end{equation}%

The entropy is therefore obtained by
\begin{eqnarray}
S &=& -  \int_{0}^{\infty} {\bf{k}}^{2}\,\mathrm{d}{\bf{k}}\phantom{a}\ln \left( 1+\chi \exp \left[ -\beta \left( E+u^{\prime }\left( \bar{n}\right) -\mu \right) \right]\right)   \notag \\
&&+ \int_{0}^{\infty} {\bf{k}}^{2}\,\mathrm{d}{\bf{k}}\phantom{a}\bar{n}_{r}\left(E+u^{\prime}\left( \bar{n}\right) -\mu \right) .  \label{eq:Entropy}
\end{eqnarray}%

The energy on the other hand is
\begin{equation}
\bar{E}= \int_{0}^{\infty} {\bf{k}}^{2}\,\mathrm{d}{\bf{k}}\phantom{a}\bar{n}_{r}E +U\left(V,\bar{n}\right) .  \label{eq:Energy}
\end{equation}%
As expected, the energy is the average of the kinetic term plus the interactions energy.  Besides, the mean occupation number can be immediately written as 
\begin{equation}
\bar{n}_{r}=\frac{1}{\exp \left[ \beta \left( E+u^{\prime }\left( \bar{n}\right) -\mu \right) \right] -1 }.
\label{eq:Mean-number-N-new}
\end{equation}%

\begin{figure*}[h!]
\centering
\begin{subfigure}{.4\textwidth}
  \centering
  \includegraphics[scale=0.35]{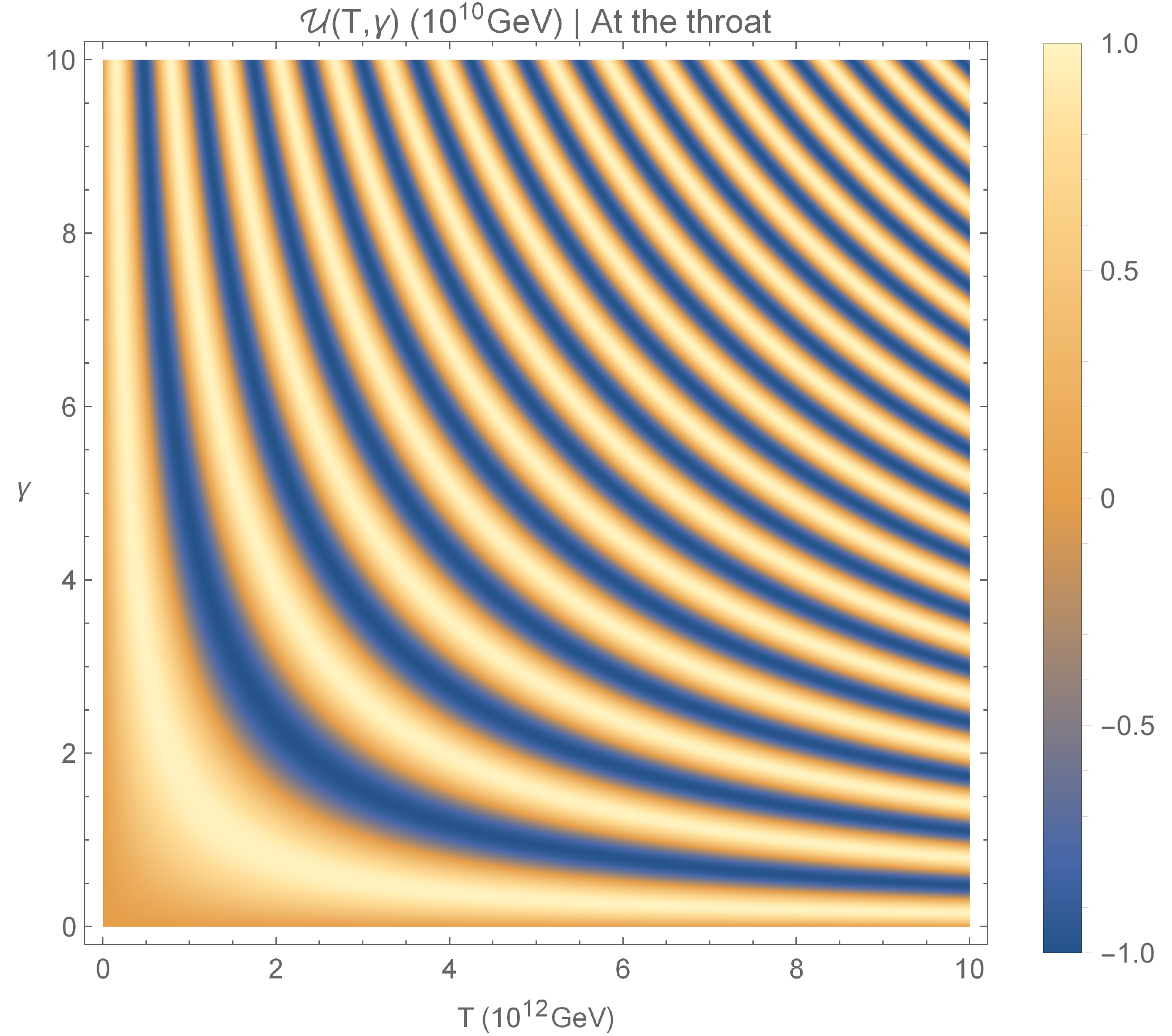}
  \caption{}
\end{subfigure}%
\begin{subfigure}{.6\textwidth}
  \centering
  \includegraphics[scale=0.35]{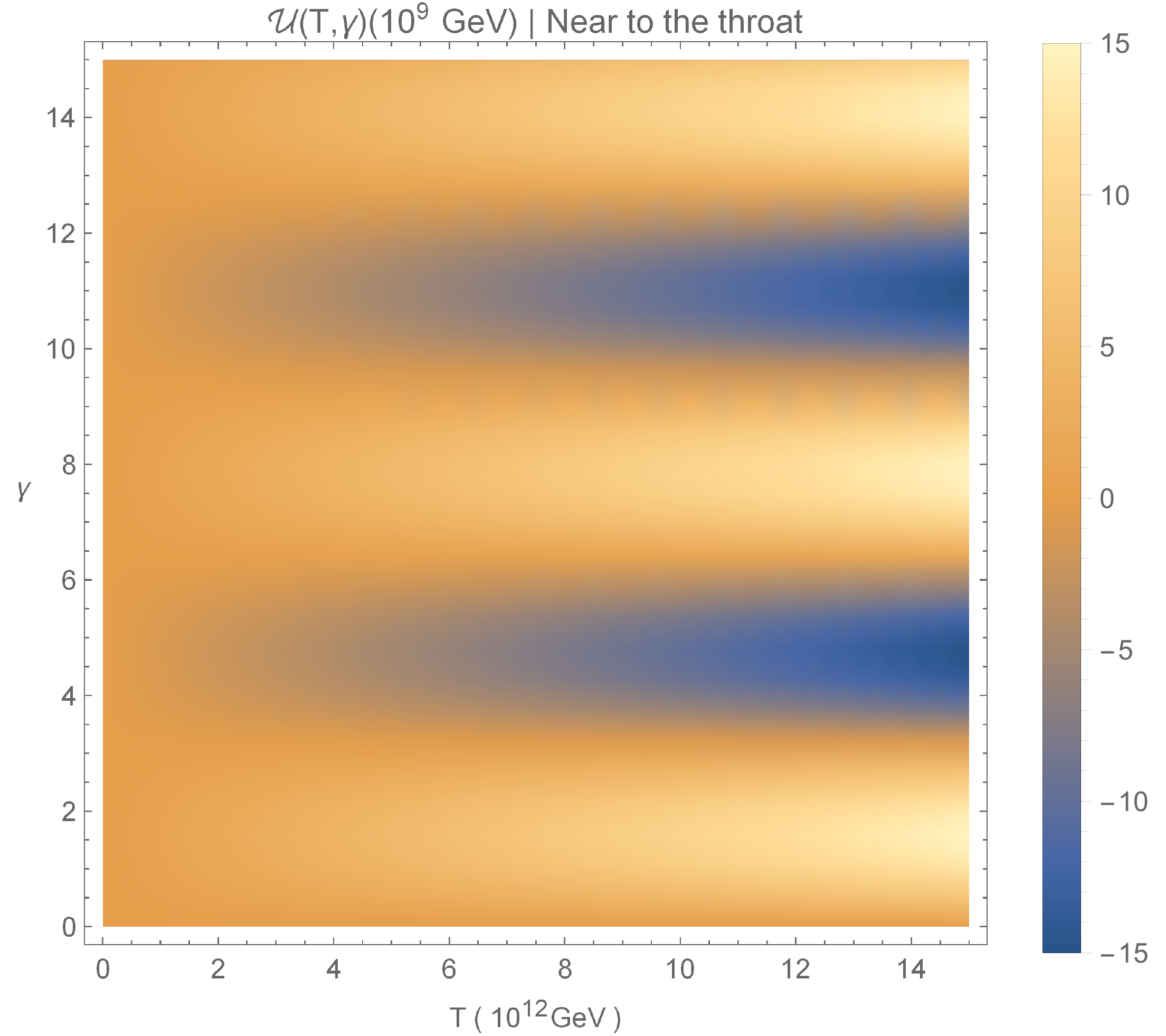}
  \caption{}
 
\end{subfigure}%
\caption{Density of energy as a function of temperature and $\gamma$.}\label{energydensity}
\end{figure*}

To enhance the clarity of the aforementioned equation, we present a plot that pertains specifically to photons within the widely recognized linear approximation. This entails retaining terms up to first order in the interaction function $u$, while approximating the density based on the established reference \cite{reis2021thermal}
\begin{equation}
    n=\frac{1}{\sqrt{2}}\left(\frac{mT}{\pi} \right)^{\frac{3}{2}} h_{\frac{3}{2}}(z).
\end{equation}
where $h_{\sigma }\left(z\right) $ is defined as
\begin{equation}
h_{\sigma }\left(z\right) =\frac{1}{\Gamma \left(
\sigma \right) }\int_{0}^{\infty }\frac{t^{\sigma -1}}{z^{-1}e^{t}-1}\mathrm{d}t.
\end{equation}%
Employing the method outlined earlier, we have generated the following plots shown in Fig. \ref{energydensity}. Notice that, since the interactions were considered, the dark energy--like behavior has vanished. This features is so relevant because it seems that, for our case, the interactions provide the ``cure" for the system.
Such an aspect will be investigate to other thermodynamic properties in an upcoming work.

\section{Final comments and perspectives}\label{IV}

In this work, we investigated the thermodynamic properties of an ideal gas on an Ellis wormhole geometry. For doing so, we derived a modified dispersion relation for massive non--interacting particles living on it. After that, all thermal quantities could be calculated in terms of the mass $M$ and the radius of the wormhole throat $r_{th}$.

The curvature of the wormhole broke the homogeneity of the spacetime, leading to position--dependent thermodynamic quantities, i.e., pressure, internal energy and heat capacity. In this sense, we investigated how these thermal properties varied with respect to parameter that controlled the radius of the wormhole throat, $\gamma$, for a given point in the spacetime.

Far from the wormhole, all the thermal variables turned out to be constants, as one could expected for an ideal fluid. However, at the wormhole throat, the thermodynamic quantities became rather dependent on parameter $\gamma$. Analogously with what happened to the wormhole radius, the pressure diverged when $\gamma\rightarrow 0$; and it went to a constant value for $\gamma\rightarrow \infty$. In the interval $2<r<3$, this thermodynamic function became negative, and, therefore, the gas behaved as a dark--energy fluid. More so, in the interval $1<\gamma<2$, the internal energy and the heat capacity became negative. This indicated that the gas underwent a transition into a rather exotic and probably unstable state.

The thermodynamic properties were compared with the simplest case: the Minkowski spacetime. An important observation was that, in the context of Minkowski spacetime, the rate at which entropy increased with temperature was faster than in the case of wormhole geometry. More so, the interactions were taken into account. In particular, we analyzed the behavior of the energy density of the system. Remarkably, interactions seemed to indicate that it ``cured" the parts where the dark energy--like behavior were present.

For future research directions, we propose investigating the behavior of fermionic and massless particles resembling photons in the context of this background geometry. This would also encompass the study of backreaction effects relevant both to these particle modes and to our current case. Additionally, examining the stability of various fluids in the vicinity of the wormhole geometry appears to be another promising avenue for investigation.


\section*{Acknowledgments}
\hspace{0.5cm}

The authors thank CNPq, FAPESQ and CAPES (Brazilian research agencies) for their financial support. Particularly, A. A. Araújo Filho was supported by Conselho Nacional de Desenvolvimento Cientíıfico e Tecnológico (CNPq) -- [200486/2022-5] and [150891/2023-7]. JF would like to thank the Fundação Cearense de Apoio ao Desenvolvimento Cient\'{i}fico e Tecnol\'{o}gico (FUNCAP) under the grant PRONEM PNE0112-00085.01.00/16 for financial support. J.E.G.Silva thanks CNPq grant  n$\textsuperscript{\underline{\scriptsize o}}$ 312356/2017-0. The authors extend their heartfelt gratitude to R. Konoplya, A.
Övgun, K. Jusufi, G. J. Olmo, and J. C. Neves for their invaluable contributions through correspondence during the development of this manuscript.


\section{Data Availability Statement}

Data Availability Statement: No Data associated in the manuscript


\bibliographystyle{ieeetr}
\bibliography{main}

\end{document}